\documentclass[twoside,12pt]{article}
\usepackage{epsfig}
\def\nin{\noindent}
\def\beq{\begin{equation}}
\def\eeq{\end{equation}}
\def\bea{\begin{eqnarray}}
\def\eea{\end{eqnarray}}
\def\enq{\end{equation}}
\def\beqa{\begin{eqnarray}}
\def\enqa{\end{eqnarray}}

\def\GeV{\nobreak\,\mbox{GeV}}

\def\lb{\label}
\def\rag{\rangle}
\def\lag{\langle}
\def\eeqa{\end{eqnarray}}

\def\pli{p^\prime}
\def\ql{{p^\prime}^2}
\def\mli{{M^\prime}^2}

\def\me#1{\langle{#1}\rangle}

\def\bra#1{\langle #1|}
\def\ket#1{| #1\rangle}
\def\qbar{\overline{q}}
\def\gs{g_{\rm s}}
\def\G{{\cal G}}

\def\xsla{\rlap{/}{x}}

\def\lp {\left( }
\def\rp {\right) }
\def\lb {\left[ }
\def\rb {\right] }

\def\bea{\begin{eqnarray}}
\def\eea{\end{eqnarray}}

\def\d {\partial }

\def\cd {\!\cdot\!}

\def\a{\alpha}

\def\G {\Gamma}

\def\L {\Lambda}
\def\m{\mu}

\def\p{\pi}

\def\r{\rho}

\def\cD {{\cal D}}

\def\bro {\mbox{\boldmath $\rho$}}
\def\bT {\mbox{\boldmath $T$}}

\def\maior{\smash{\mathop{>}\limits_{\raise4pt\hbox{$\scriptstyle \sim$}}}}
\def\menor{\smash{\mathop{<}\limits_{\raise4pt\hbox{$\scriptstyle \sim$}}}}

\begin{document}

\title{Charm couplings and form factors in  QCD sum rules}


\author{M.E. Bracco$^1$, M. Chiapparini$^2$, F.S. Navarra$^3$, M. Nielsen$^3$ \\
$^1$Faculdade de Tecnologia, Universidade do Estado do Rio de 
Janeiro, \\
Rod. Presidente Dutra Km 298, Polo Industrial, 27537-000, Resende, RJ, Brazil \\
$^2$Instituto de F\'{\i}sica, Universidade do Estado do Rio de Janeiro, \\
Rua S\~ao Francisco Xavier 524, 20550-900 Rio de Janeiro, RJ, Brazil \\
$^3$Instituto de F\'{\i}sica, Universidade de S\~{a}o Paulo, \\
C.P. 66318, 05389-970 S\~{a}o Paulo, SP, Brazil}

\maketitle

\begin{abstract}
We review the  calculations of  form factors  and  coupling constants in  vertices 
with charm mesons   in the framework of QCD sum rules. We first discuss the motivation 
for this 
work, describing possible applications of these form factors to heavy ion collisions 
and to B decays. We then present an introduction to the method of QCD sum rules and 
describe how to work with the three-point function. We give special attention to the 
procedure employed to extrapolate results obtained in the deep euclidean region to the 
poles of the particles, located in the time-like region. We present a table of 
ready-to-use parametrizations of all the form factors, which are relevant for the 
processes mentioned in the introduction. We discuss the uncertainties in our results. 
We also give the coupling constants and compare 
them with estimates obtained with other methods. Finally we apply our results to the 
calculation of  the cross section of the reaction $J/\psi + \pi \rightarrow 
D  + \overline{D^*}$. 
\end{abstract}

\maketitle

\vspace{0.5cm}

\begin{center}
{\sl Prepared for publication in Prog. Part. Nucl. Phys.}
\end{center}

\newpage

\tableofcontents

\newpage

\section{Introduction}
\label{sec_intro}

When we write the  high energy electron-proton scattering cross section using the  
Feynman rules of quantum electrodynamics (QED), we do not know how to write the 
interaction between the photon and the 
proton when the latter is an extended and composite object. In order to organize and parametrize 
our ignorance making use of general principles, such as charge conservation, we introduce
functions of the involved kinematical variables, which are called structure functions. 
After convenient redefinitions these functions become form factors. The final cross 
sections, written in terms of form factors are then adjusted to data and, in this way, 
the form factors are measured. In a particular reference frame, their Fourier transform 
gives the spacial distribution of charged matter in the proton. This procedure gives a 
nice ``picture'' of the proton in space and the most accurate description of its ``form''.  
In lower energy experiments, where 
also the four momentum transfer ($q^2$) is low,  it was possible to determine 
the electromagnetic form factor  (and the charge radius) of the nucleon \cite{texto}.

QED was also applied  to  deep inelastic  electron-proton scattering with the assumption 
that the proton is composed by pointlike constituent fermions. In this case the structure 
functions turned into the parton distribution functions and their precise measurement gave
reality to quarks and gluons and illuminated a new world within  the proton. In 
higher energies experiments and very large values of $q^2$ a very different 
picture of the nucleon emerged, in which it is made of pointlike particles, 
the quarks. From these  observations one may conclude that, when probing the 
nucleon, nearly on-shell photons ($q^2 \simeq 0$) recognize  sizes whereas 
highly off-shell photons ($q^2 \ll 0$) do not \cite{texto}.  This statement is supported 
by the phenomenologically very successful vector meson dominance hypothesis, 
according to which real photons are with a large probability converted to 
vector mesons (which are extended objects) and then interact with the nucleon \cite{vmd}. 
HERA data on electron-proton reactions can be well understood 
introducing a ``transverse radius of the photon'', parametrized as
\begin{equation}
r_{\gamma} \simeq \frac{1}{\sqrt{Q^2 + m^2}} , 
\end{equation}
where $Q^2 = - q^2$ and $m$ is the mass of the vector meson considered. This 
empirical formula tells us that for $Q^2 \rightarrow \infty$ the photon is 
pointlike  and ``resolves'' the nucleon target, i.e., identifies its 
pointlike constituents and does not ``see''  the size of the nucleon.
Moreover, this formula indicates that for $Q^2 \simeq 0$ and for light mesons
(like the $\rho^0$) the photon has appreciable transverse radius  and therefore 
also identifies the global nucleon extension. Finally, in the above formula we 
may have a heavy vector meson ($J/\psi$ or $\Upsilon$) which will, either 
real or virtual, resolve the nucleon into pointlike constituents. This feature nicely 
explains why the $J/\psi$ photoproduction cross section grows 
pronouncedly with the (photon-proton) energy whereas the $\rho$ cross section
grows very slowly. In the former case the compact $J/\psi$ interacts with 
the small $x$ gluons in the protons, which have a fastly growing population. 
In the latter case, the $\rho$ identifies the global and slowly growing 
geometrical size of the proton.

After this glorious history as an important tool in the QED machinery, form factors were 
calculated in other contexts. In particular they were introduced in the development of 
effective theories to study interactions of charm particles. In this context, an interesting 
question is:  which are the form factors of the charm  mesons when probed by light 
particles such as photons, pions and $\rho$ mesons?  
Apart from their intrinsic value as fundamental knowledge about nature, the answer to this 
question has  immediate applications in hadron physics.

Vertices with three mesons where at least two of them carry charm appear in theories of
the charm meson interactions. This kind of theory started to become popular in the late 
nineties \cite{mamu}, during the analysis of the CERN-SPS data on charmonium production in 
heavy ion collisions. At that time it was believed that $J/\psi$ suppression was a 
good signature of quark gluon plasma formation (QGP) \cite{matsui}. However a careful 
evaluation 
of the background was needed, in order to isolate the signal. In this case the background was 
the charmonium absorption by light mesons within the hadronic fireball formed at the 
late stage of these collisions. Since the center of mass energy of these collisions was  
of the order of magnitude of the temperature, i.e. $\simeq 100-200$ MeV, the interaction 
regime was clearly non-perturbative and  the best tools were the effective Lagrangian 
models with charm mesons \cite{mamu,galera}. From these Lagrangians one can derive the 
Feynman rules 
and compute scattering amplitudes. In order to avoid infinities one can introduce form 
factors in the vertices. 
In phenomenological applications, these form factors contain a parameter $\Lambda$ which 
plays the 
role of a cut-off.  In fact, even diagrams which give finite contributions to the cross 
sections must 
contain form factors. Otherwise the obtained cross sections are unacceptably large. After 
the 
introduction of form factors the results for the charmonium interaction cross section  
become quite 
reasonable. However these results depend too strongly on the choice of the cut-off 
parameter. 
In QCD sum rules (QCDSR) we can calculate these form factors from first principles, 
eliminating the  freedom in the choice of parameters.

At this point one might argue that this program is futile for at least two reasons. In 
first place one
might say that one can compute cross sections such as, for example,
 $J/\psi + \pi \rightarrow D +  
\overline{D}$, directly from QCDSR \cite{nos_cross} and it is not 
necessary  to compute form factors. This statement is in principle 
correct. In practice however, the direct calculation of cross sections with QCDSR requires the use of the 
four-point function, which is much less precise than the three and two-point functions. It is not clear, 
for example, whether one should perform one, two or even three Borel transforms. In second place one might 
say that it is more promising to develop a chiral perturbation theory for these interactions, where one 
eliminates form factors \cite{chiral}. This may indeed be the case, but for now this kind of theory is in a very 
preliminary stage. Moreover, in effective theories one needs the coupling constants. In the case of the charm 
three meson  vertices these couplings (with the exception of the $D^* D \pi$ coupling constant) are not 
measurable. They can be calculated in QCDSR, as a by-product of the calculation of form factors. 

The experimental study of interactions of charmed mesons with nucleons and also with light mesons will be one of 
the main topics of the  scientific program of the PANDA and CBM experiments at the future FAIR facility at GSI 
\cite{fair}. 

\begin{figure}[h]
\begin{center}
\epsfxsize=9cm
\leavevmode
\hbox{\epsffile{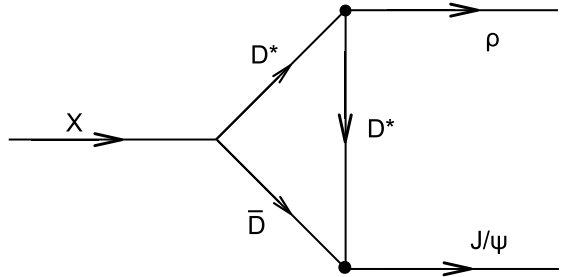}}
\end{center}
\protect\caption{Final state interactions with charm meson loops.}
\label{figchina}
\end{figure}

So far we have been emphasizing the usefulness of charm form factors in the calculation of charmonium interaction 
cross sections.  However, this is just one of the several applications of these form factors. Another context where
they are needed is in heavy meson decays. Since 2003, due to the precise measurements of $B$ decays  performed by 
BELLE, BES and BABAR, charm form factors  gained a new relevance. In $B$ decays new particles have been observed, 
such as the $D_{sJ} (2317)$ and the $X(3872)$ \cite{rev, nora}. These particles decay very 
often  into  an intermediate 
two body state, which then undergoes final state interactions, with the exchange of one or more 
virtual mesons \cite{china}.  As an example of  specific situation we may consider the decay 
$X(3872) \, \rightarrow \, J/\psi \, + \, \rho$. This decay may proceed in two steps. First the $X$ decays 
into a $D$ - $D^*$ intermediate state and then these two particles exchange a $D^*$ producing the final $J/\psi$ and $\rho$ 
\cite{lzz}.  This process is illustrated in Fig. \ref{figchina}.  In order to compute 
the effect of these interactions in the final decay rate we need the $ \rho  D^*  D^*  $ and   
$ J/\psi   D^*  D  $ form factors. 

Besides the decays, charm form factors may be needed to understand the structure of these newly observed charm states. In 
\cite{molina1,molina2,molina3} it has been suggested that some of these states are dinamically generated resonances from 
the vector-vector mesonic interaction. Here we have  box diagrams with virtual meson exchanges and vertices which need 
form factors. 

In this article  we review our works on charm form factors 
\cite{00,01,02,02r,05,05r,05i,08,10} calculated with  QCD  sum rules 
\cite{rev,svz,rry,nar}. In doing so we compare our method with other approaches, pointing out 
their virtues and shortcomings. The text is organized as follows.  In Section 
\ref{sec_qcdsr} we present a brief  
introduction to the QCDSR method, 
where we emphazise the main concepts and the strategies employed to calculate masses, 
decay constants, form factors 
and coupling constants. This introduction is intentionally  abstract and ``clean''.  
The subsequent Section 
\ref{sec_eval} illustrates how to calculate numerically all desired quantitities. 
This is where the limitations of 
the method become clear and also where we show how to circunvent them. 
In this section we discuss the quality of 
numerical results and establish certain conditions which must be fulfilled  by 
reliable sum  rules.  After having 
defined the method and the quality criteria, in Section \ref{sec_form} we present 
a sample of results obtained for 
the form factors. We give special attention to one particular form 
factor and the respective vertex coupling constant: 
$D^* D \pi$. This vertex is special because is the only one directly accessible 
to experiments. Its  coupling constant 
was measured and  several calculations were performed, one of them with lattice 
QCD techniques. The $D^* D \pi$ vertex 
provides the precision test for the different ways of computing the coupling constant.
 A careful analysis of this vertex 
resulted in the formulation of a complementary technique  
(the evaluation of meson loops in the vertex) used to improve the 
quality of the QCDSR calculations.   In Section \ref{sec_res} we show the full 
compilation of all the already obtained 
charm form factors, with the corresponding  parametrizations, 
which are ready to  be easily employed in phenomenological 
analyses of  vertices with charm mesons.  We also present 
a quantitative study of the uncertainties in our results.  
In Section \ref{sec_disc} we perform a comparison with the results obtained with 
other approaches.  	
In Section \ref{sec_psi} we show an application of 
the charm form factors to the study of $J/\psi$ production and absorption by light mesons. 
Finally, in Section  \ref{sec_conc} we present  a summary of our results.

\section{QCD Sum Rules}
\label{sec_qcdsr}

QCD sum rules have been discussed in many reviews \cite{rev,svz,rry,nar}
emphasizing various aspects of the method. QCDSR are 
a powerful tool to extract qualitative and quantitative information 
about hadron properties. In this approach we start with a correlation
function that is constructed in terms of hadronic currents, which are chosen so as
to have the quantum numbers of the hadrons in question.
The basic idea of the formalism is to approach the bound state problem in 
QCD from short distances and move to larger distances, including 
non-perturbative effects and using some approximate procedure to extract 
hadronic masses and couplings. Although we are mostly concerned with the three-point 
correlation function, sometimes we  also need  two-point correlation  functions. 
For this reason and also in order to illustrate some basic procedures in QCDSR we  
devote the next subsections to the discussion of two-point correlators. 

\subsection{The two-point correlation functions}

A generic two-point correlation function is usually written as: 
\beq
\Pi(q)\equiv i\int d^4 x\, e^{iq\cdot x}
\lag{0}| T [j(x)j^\dagger(0)]|0\rag\ ,
\label{sr-cor}
\enq
where $j(x)$ is a current with the quantum numbers of the hadron which we want to
study.

In QCDSR we  assume that the  correlation functions  may be written at both the  
quark and the hadron levels. Identifying the hadronic representation with the 
corresponding representation in terms of quarks and gluons, we obtain the sum 
rule, from  which we can determine hadron properties. 
In the QCD side we proceed with the calculation of the correlation function using 
the operator product expansion (OPE), which is formulated with Wilson coefficients 
and local operators in terms of the non-perturbative structure of the QCD vacuum.
In order to apply this method to the correlation function (\ref{sr-cor}), we 
have to expand the product of two currents in a series of local operators:
\beq
\Pi^{OPE}(q^2)= i\int d^4 x\, e^{iq\cdot x}
\lag{0}| T[j(x)j^\dagger(0)|0\rag\ =\sum_nC_n(q^2) 
\langle 0 |  \hat{O}_n  | 0 \rangle  ,
\label{cope}
\enq
where the set $\{\hat{O}_n\}$ includes all local gauge invariant
operators expressible in terms of light quark and gluon fields. 
By construction the coefficients $C_n(Q^2)~(Q^2=-q^2)$ in Eq.~(\ref{cope})  
include only  short-distance effects. Therefore, they can be evaluated
perturbatively. Non-perturbative long-distance effects are contained
only in the local operators. In the OPE expasion, the operators are
ordered according to their dimension $n$. The lowest-dimension
operator with $n=0$ is the unit operator associated with the perturbative
contribution: $C_0(Q^2)=\Pi^{per}(Q^2)$, $\hat{O}_0=1$. The QCD
vacuum fields are represented in (\ref{cope}) in the  form of the so
called vacuum condensates. The lowest dimension condensates are the
quark condensate of dimension three: 
$\langle 0 | \hat{O}_3  | 0 \rangle =\langle\bar{q}q\rangle$,
and the gluon condensate of dimension four: 
$\langle 0 |  \hat{O}_4  | 0 \rangle = \langle g^2 G^2\rangle$. 
The contributions of higher dimension
condensates are suppressed by large powers of $\Lambda_{QCD}^2/Q^2$,
where $\Lambda_{QCD}$ is the typical long-distance scale. Therefore,
even at intermediate $Q^2\sim1\GeV^2$, the expansion in Eq.~(\ref{cope})
can be safely truncated after a few terms.

\subsection{The spectral density}

The hadronic (or ``phenomenological'')  representation of the correlation 
function in Eq.~(\ref{sr-cor}) can be written as  a dispersion relation:
\beq
\Pi^{phen}(q^2)=-\int ds\, {\rho(s)\over q^2-s+i\epsilon}\,+\,\cdots\,,
\label{disp}
\enq
The dots in the above equation  represent subtraction terms.
The spectral density is given by  the imaginary part of the correlation
function:
\beq
\rho(s)={1\over\pi}Im[\Pi(s)]\;.
\enq
The evaluation of  $\rho(s)$ is simpler than the
evaluation of the correlation function itself, and the knowledge of
$\rho(s)$ allows one to recover the whole function $\Pi(q^2)$ through the
integral in Eq.~(\ref{disp}).

The current $j~(j^\dagger)$  appearing in (\ref{sr-cor}) and (\ref{cope})   
is an operator that annihilates (creates) all hadronic states that have the 
same quantum numbers as $j$. Consequently, $\Pi(q^2)$ contains information not only
about the low mass hadron of interest, but also about all excited states with the
same quantum numbers. When comparing a set of hadrons with the same quantum 
numbers, the lowest resonance is often fairly narrow, whereas higher mass states 
are broader. We can therefore parametrize the spectral density as a single sharp
pole, representing the lowest resonance of mass $m$, plus a smooth continuum, 
representing higher mass states:
\beq
\rho(s)=\lambda^2\delta(s-m^2) +\rho^{cont}(s)  ,  
\label{den}
\enq
In the above Equation the parameter $\lambda$ represents the coupling of the 
current to the low mass hadron  $H$:
\beq
\lag 0 |
j|H\rag =\lambda.
\label{cou}
\enq
For simplicity, we often assume that the continuum contribution to the
spectral density, $\rho^{cont}(s)$ in Eq.~(\ref{den}), vanishes bellow a
certain continuum threshold $s_0$. In order to keep the number of parameters as 
small as possible, we assume  that above $s_0$ the spectral density is given by 
the result obtained with the OPE. This idea is implemented by the Ansatz 
proposed in \cite{ioffe81}:
\beq
\rho^{cont}(s)=\rho^{OPE}(s)\Theta(s-s_0) .
\label{ansacom}
\enq
This is called quark-hadron duality.

\subsection{The mass sum rule} 

The sum rule is obtained from the matching of the two descriptions of the 
correlator:
\beq
\Pi^{phen}(Q^2) =  \Pi^{OPE}(Q^2)\;.
\enq
However, such a matching is not yet practical. The phenomenological description 
is significantly dominated by the lowest pole only for sufficiently small $Q^2$, 
or even better, timelike $q^2$ near the pole. On the other hand, the OPE side is
only valid at a sufficiently large spacelike $Q^2$. In order to improve the overlap 
between the two sides of the sum rule, we apply the Borel transform:
%
\begin{equation}
{\cal B}_{M^2}[\Pi(q^2)]=
\lim_{\stackrel{\scriptstyle -q^2,n\rightarrow\infty}
{\scriptstyle -q^2/n=M^2}}
{(-q^2)^{n+1}\over n!}\left(d\over d q^2\right)^n
\Pi(q^2)\;.
\label{borel}
\end{equation}
It is interesting to notice that:
\beq
{\cal B}_{M^2}\left[{q^2}^n\right]=0\, ,  \,\,\,\,  n>0\;.
\label{po}
\enq
This means that all the subtractions terms in Eq.~(\ref{disp}) are eliminated by 
the Borel transform. Another important feature of the Borel tranform is the fact
that:
\beq
{\cal B}_{M^2}\left[\frac{1}{(m^2-q^2)^n}\right] = \frac{1}{(n-1)!}
\frac{e^{-m^2/M^2}}{(M^2)^{n-1}}\;, \,\,\,\,  n>0\;. 
\label{fra}
\enq 
Therefore, the Borel transform exponentially suppresses the contribution from 
excited resonances and continuum states in the phenomenological side. In the OPE 
side the Borel transform suppresses the contribution from higher dimension 
condensates by a factorial term,  improving  the OPE convergence.
After making a Borel transform on both sides of the sum rule, and
transferring the continuum contribution to the OPE side, the sum rule
can be written as
\beq \lambda^2e^{-m^2/M^2}=\int_{s_{min}}^{s_0}ds~
e^{-s/M^2}~\rho^{OPE}(s)\;. 
\label{sr} 
\enq
The mass of the low-lying state, $m$, can be determined by taking the derivative 
of Eq.~(\ref{sr}) with respect to $1/M^2$, and dividing the result by Eq.~(\ref{sr}). 
This gives:
\beq
m^2={\int_{s_{min}}^{s_0}ds ~e^{-s/M^2}~s~\rho^{OPE}(s)\over\int_{s_{min}}^{
s_0}ds ~e^{-s/M^2}~\rho^{OPE}(s)}\;.
\label{m2}
\enq
Since in the evaluation of both sides of the sum rule we have to make
approximations, the value extracted from Eq.~(\ref{sr}) will be a function
of $M^2$. The Borel window is defined as the range of values of $M^2$ where
the two sides of the sum rule have a good overlap and, therefore, information 
on the lowest resonance can be extracted.
In general the Borel window is determined by imposing two different criteria:
the minimum value of the Borel mass is fixed by requiring the convergence of 
the OPE and the maximum value of the Borel mass is determined by imposing the 
condition that the pole contribution must be bigger than the continuum 
contribution \cite{nos_criterio}.


\subsection{The three-point function}

The three-point function associated with a generic  vertex of three mesons $M_1$, $M_2$ and $M_3$ 
is given by
\beq
\Gamma (p,\pli) = \int d^4x \, d^4y \;\; e^{i\pli\cdot x} 
\, e^{-i(\pli-p)\cdot y}  \langle 0|T\{j_{3}(x) j_{2}^{\dagger}(y) 
 j_{1}^{\dagger}(0)\}|0\rangle\,  ,
\label{corr} 
\enq
where the current $j_i$ represents states with the quantum numbers of the meson $i$. As in the previous Section, 
the correlation function is evaluated in two ways. In the first one, we consider that the currents are composed by
quarks and write them in terms of their flavor and color content with the correct quantum numbers. This is the QCD 
description of the correlator and  it   gives 
rise to the QCD side (or OPE side) of the sum rule. 
In the second way, we write the correlation function in 
terms of matrix elements of hadronic states which can be extracted from experiment, or 
calculated with lattice QCD or estimated with effective Lagrangians.  
In this last approach we 
never talk about quarks and use all the available experimental information concerning the masses and decay 
properties of the relevant mesons. This is the hadronic description of the correlator and is called the 
phenomenologial side of the sum rule.  After studying both sides  separately, we identify one description with 
the other and write the sum rule.

\subsection{The OPE side}

\begin{figure}[h]
\begin{center}
\epsfxsize=10cm
\leavevmode
\hbox{\epsffile{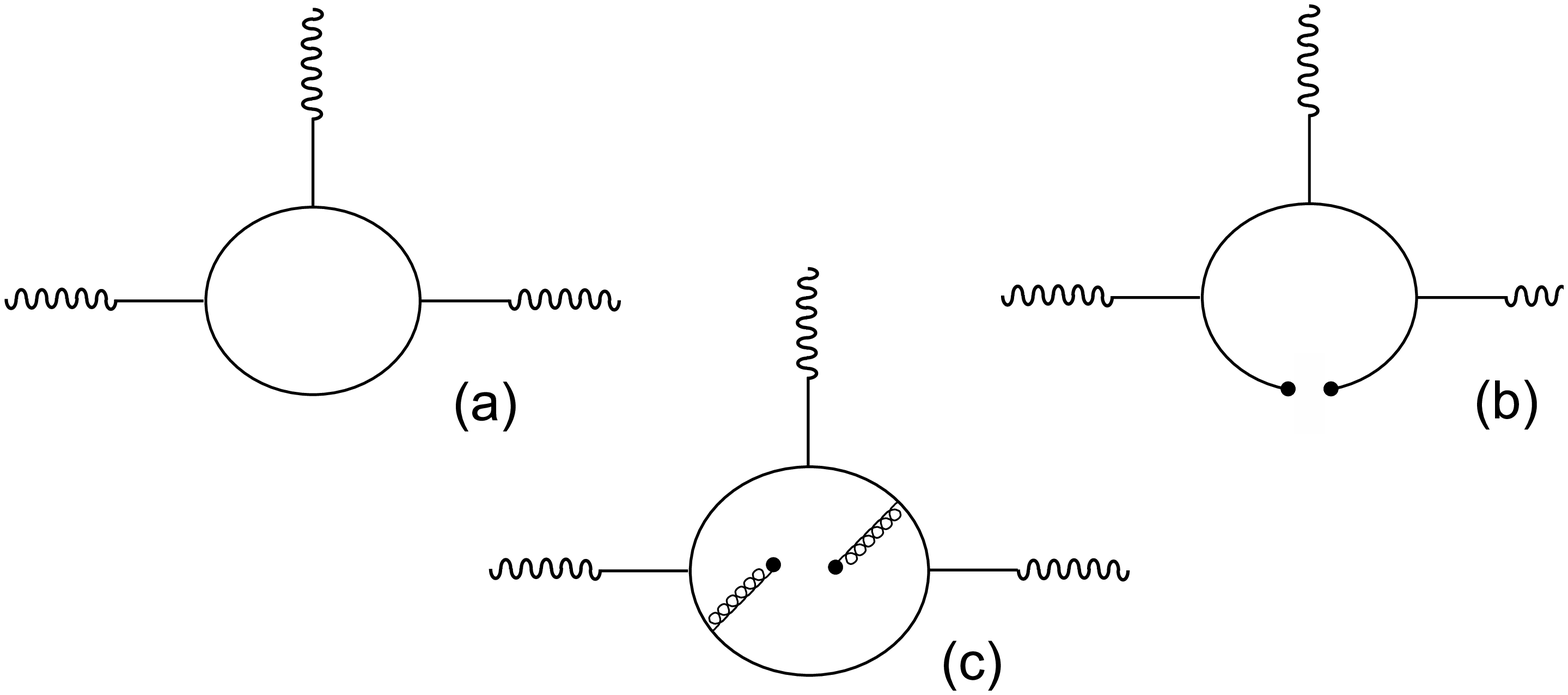}}
\end{center}
\protect\caption{Diagrammatic representation of the first terms of the OPE series for a 
three-point funtion. (a) The perturbative term. (b) The quark condensate. (c) The 
gluon condensate.}
\label{opegraf}
\end{figure}

We start with the current,  which has the general form:
\beq
j_i = \bar{q}  \, \Lambda \, q  ,
\label{curr}
\eeq
with $\Lambda = 1, \gamma_{\mu}, \gamma_5, \gamma_{\mu} \gamma_5$ for a scalar, vector, 
pseudo-scalar 
and axial-vector meson, respectively.  $q$ is the quark spinor field. Since the currents may carry Lorentz indices, 
so will the vertex function $\Gamma$. When we insert the three currents into Eq. (\ref{corr}) we get the 
vacuum expectation value of the $T$ product of six quark fields multiplied by Dirac matrices in different points.
In this expression we apply Wick's theorem obtaining a series of terms, each of which being the product of 
contractions (propagators) times a vacuum expectation value of normal ordered operators taken in the QCD vacuum. 
These latter, in the local approximation, are  the QCD condensates. This series is 
precisely (\ref{cope}) and its first terms are diagramatically depicted in 
Fig. \ref{opegraf}. The first and leading one is  the perturbative term in Fig. 2a. Figs. 2b and 2c show examples of 
terms with the quark condensate  and with the gluon condensate respectively. 

The evaluation of (\ref{corr}) in lowest order  leads  to a loop diagram with three quark propagators. 
The first corrections to the simple bubble (called the 
perturbative contribution) come from diagrams where a quark line ``dives'' into the vacuum and emerges from it.
This contribution to the correlators appears multiplied by the quark condensate 
$\langle \bar{q} q \rangle$. Higher order 
corrections represent the interactions of the quark lines exchanging gluons among themselves 
($\alpha_s$ corrections) and with the vacuum (gluon condensates and quark-gluon mixed condensates).  An equivalent 
description is obtained performing first the contractions in (\ref{corr}), 
obtaining three quark propagators and then performing an operator product expansion (OPE) of these propagators. 
For our purposes it is enough to consider only the first few terms of this expansion:
\beqa
S_{ab}(x)&=&\bra{0} T[q_a(x)\overline{q}_b(0)]\ket{0}=
{i\delta_{ab}\over2\pi^2x^4}\xsla - {m_q\delta_{ab}\over4\pi^2x^2} 
-{\delta_{ab}\over12}\me{\qbar q} + {i\delta_{ab}\over48}m_q\me{\qbar q}\xsla
\nonumber\\
&-&{i\over32\pi^2x^2}T^A_{ab} \gs G^A_{\mu\nu}(\xsla\sigma^{\mu\nu}+\sigma^{\mu\nu}\xsla)
-{m_q\over32\pi^2}T^A_{ab}\gs G^A_{\mu\nu}\sigma^{\mu\nu}\ln(-x^2) .
\label{ope_propag}
\eeqa
In the above expression $x$ is the four-vector which defines the separation between the two quark fields, 
$m_q$ is the quark mass, $T^A_{ab}$ are the SU(3) generators, $G^A_{\mu\nu}$ is the gluon field tensor and 
$\sigma^{\mu\nu}$ is the anticommutator of the Dirac  matrices. In (\ref{ope_propag}) the first two terms 
are related to Fig. 2a, the next two terms are related to Fig. 2b and the last two are related to Fig. 2c. 
After the contractions and the OPE expansion we arrive at an expression for $\Gamma$ which has the generic form:
\beq
\Gamma_{OPE} = \sum_j F_j^{OPE}(p^2,{\pli}^2,q^2) \, L_j ,
\label{struc}
\eeq
where $F_j^{OPE}(p^2,{\pli}^2,q^2)$ are invariant functions of the momenta and $L_j$ are 
the structures, i.e., products of
Dirac matrices, the metric tensor and the four momenta, carrying Lorentz indices. 
For each one of the invariant amplitudes appearing in Eq.(\ref{struc}), we 
can write a double dispersion relation over the virtualities $p^2$ and  ${\pli}^2$:
\begin{equation}
F_i^{OPE} =-\frac{1}{4 \pi^2}\int_{s_{min}}^\infty ds
\int_{u_{min}}^\infty du \:\frac{\rho_i^{OPE}(s,u,q^2)}{(s-p^2)(u-{\pli}^2)} , 
\label{dis}
\end{equation}
where $\rho_i^{OPE}(s,u,q^2)$ is the double discontinuity of the amplitude
$F_i^{OPE}(p^2,{\pli}^2,q^2)$ and can be calculated using the Cutkosky's rules. 
We can work with any structure appearing in
Eq.(\ref{struc}), but we must choose those which have less ambiguities in 
the QCD sum rules approach, which means among other things, a weak influence from the 
higher dimension 
condensates. The invariant amplitudes and thus the double discontinuities $\rho$,  
receive contributions from all terms in the OPE. The first one (and dominating) of 
these contributions must come from  the perturbative term.

\subsection{The phenomenological side}

\label{phenside}

For the phenomenological side we formally repeat the steps mentioned above and  find that, 
as before, the correlation function can be written as a sum of contributions with different 
tensor structures, i.e. :
\beq
\Gamma_{phen} = \sum_j F_j^{phen}(p^2,{\pli}^2,q^2) \, L_j , 
\label{strucn}
\eeq
where $F_j^{phen}(p^2,{\pli}^2,q^2)$ are invariant functions of the momenta and $L_j$ are the 
structures. For each one of the invariant amplitudes appearing in Eq.(\ref{strucn}), we 
can write a double dispersion relation over the virtualities $p^2$ and 
${\pli}^2$, holding $Q^2= -q^2$ fixed:
\begin{equation}
F_i^{phen} = -\frac{1}{ 4 \pi^2}\int_{s_{min}}^\infty ds
\int_{u_{min}}^\infty du \:\frac{\rho_i^{phen}(s,u,Q^2)}{(s-p^2)(u-{\pli}^2)} , 
\label{disn}
\end{equation}
where $\rho_i^{phen}(s,u,Q^2)$ is the double discontinuity of the amplitude
$F_i^{phen}(p^2,{\pli}^2,Q^2)$. The function $\rho$ can be generically written as: 
\beqa
\rho_i^{phen}(s,u,Q^2) &=& a \, \delta(s - m_1^2) \, \delta(u - m_2^2) \,  +
                         b \, \delta(s - m_1^2) \, \theta(u - u_0)   \,  + \nonumber  \\
                       &&  c \, \delta(u - m_2^2) \, \theta(s - s_0)   \,  + 
\rho^{cont}_i (s,u,Q^2) \, \theta(s - s_0) \, \theta(u - u_0) , 
\label{doubdiscn} 
\eeqa
where $s_0$ and $u_0$ are the continuum threshold parameters. In the above equation 
the first term represents a kinematical situation where the mesons $M_1$ and $M_2$ 
(with masses $m_1$ and $m_2$ respectively) are in  
the ground state and $M_3$ is off-shell, having an arbitrary Euclidean four momentum squared
$Q^2$.   The second term represents a situation where the meson $M_1$ is on-shell but the
groundstate of the meson $M_2$ is absent in the vertex, which contains only its excitations starting at $u_0$.
The third term represents the inverse situation, where  $M_2$ is on-shell and it is 
accompanied in the vertex by the excitations of $M_1$ starting at $s_0$. Finally, the 
last term represents the excitations of  $M_1$ and $M_2$, which start at 
$s_0$ and $u_0$ respectively. After a double Borel transform in the variables $p^2$ and 
${\pli}^2$ the second and third terms are exponentially suppressed. One can then safely 
assume quark-hadron duality and parametrize the continuum by the double discontinuity of 
the theoretical part: $\rho_i^{cont}(s,u,Q^2)=\rho_i^{OPE}(s,u,Q^2)$. 
This point is thoughtfully discussed in \cite{ioffe_smilga}.
Inserting the above expression into (\ref{disn}) (and neglecting the suppressed terms) 
we have:
\beq
F_i^{phen} = \Lambda_i^{phen} - \frac{1}{4 \pi^2}\int_{s_{0}}^\infty ds
\int_{u_{0}}^\infty du \:\frac{\rho_i^{OPE}(s,u,Q^2)}{(s-p^2)(u-{\pli}^2)} , 
\label{strucfin}
\eeq
where $\Lambda_i^{phen}$ refers to the ground state (or pole) contribution in the structure $i$. In order to calculate $\Lambda_i^{phen}$ we go back to (\ref{corr}) and 
insert complete sets of hadronic states. After some algebra \cite{derafael}  we obtain:
\beq
\Lambda_i^{phen} = \frac{f_{M_{1}} f_{M_{2}} f_{M_{3}} \langle M_1 M_2 M_3 
\rangle_i}  {(p^2 - m_1^2)({\pli}^2 -m_2^2)(q^2 -m_3^2)} . 
\label{corrphen}
\eeq
The meson decay constants appearing in the above equation are  defined by the following  
matrix elements:
\beq
\langle 0|j^{V}_{\mu}| V \rangle =  m_{V} f_{V} \, \epsilon_{\mu} ,
\label{fmv}
\eeq
\beq
\langle 0|j^{A}_{\mu}| A \rangle =  m_{A} f_{A}  \, \epsilon_{\mu} ,
\label{fma}
\eeq
\beq
\langle 0|j_{5}| P \rangle=  \frac{m^2_P}{m_q}  f_{P} ,
\label{fmp}
\eeq
and
\beq
\langle 0|j^{\mu}_{A}| P \rangle= i  f_{P} \, p^{\mu} ,
\label{fmpa}
\eeq
for vector, axial, pseudoscalar and axial currents respectively. 
In the above expressions $\epsilon^{\mu}$   and  
$p^{\mu}$ are  the relevant  polarization vector  
and  four momentum of the involved mesons resepectively.  
Eq. (\ref{fmpa}) refers to the case where a pseudoscalar 
meson is represented by an axial current, and $m_q$ in Eq.~(\ref{fmp}) is
the heaviest quark in the pseudoscalar meson $P$.

Since the higher states have been considered in the second term of (\ref{strucfin}), 
the amplitude $\langle M_1 M_2 M_3 \rangle$ appearing in (\ref{corrphen})  refers only to 
the ground states of the mesons $M_1$, $M_2$ and $M_3$ and can be calculated with the help 
of an effective Lagrangian of the type:
\beq
\mathcal{L}_{M_1 M_2 M_3}= g_{M_1 M_2 M_3}  \Big( \eta \,  
\Delta M_1 \, \Delta M_2 \, M_3 \, + \,  hc \Big) , 
\label{lagr}
\eeq
where $\eta = i$ or  $\varepsilon_{\mu \nu \alpha \beta}$,  $\Delta = 1$ or $\partial_{\mu}$,  
$hc$ stands for Hermitian conjugates and $g_{M_1 M_2 M_3}$ is the coupling constant. 
The choices implied in $\eta$ and $\Delta$ depend on the specific combination of mesons.

\subsection{The sum rule}

The sum rule is obtained from the identity:  
\beq
{F_i}^{phen}(p,\pli,q) = {F_i}^{OPE}(p,\pli,q) ,
\eeq
that gives
\beq
\frac{f_{M_{1}} f_{M_{2}} f_{M_{3}} \langle M_1 M_2 M_3 \rangle_i}  
{(p^2 - m_1^2)({\pli}^2 -m_2^2)(q^2 -m_3^2)}=-\frac{1}{ 4 \pi^2}
\int_{s_{min}}^{s_0} ds
\int_{u_{min}}^{u_0} du \:\frac{\rho_i^{OPE}(s,u,Q^2)}{(s-p^2)(u-{\pli}^2)},
\label{sumrule}
\eeq
 where we have  used (\ref{dis}), (\ref{strucfin}) and (\ref{corrphen})  and 
have transfered the dispersion integral 
in the phenomenological side to the OPE side.  Notice that the integration limits 
are now finite. 
In order to improve the matching between the two sides of the sum rule and also to 
suppress the 
pole-continuum transitions \cite{ioffe_smilga} we perform a double Borel 
transform (\ref{borel}) in the variables 
$P^2=-p^2\rightarrow M^2$ and $P'^2=-{\pli}^2\rightarrow M'^2$, 
on both sides:
\beq
\frac{f_{M_{1}} f_{M_{2}} f_{M_{3}} \langle M_1 M_2 M_3 \rangle_i}  
{(Q^2 +m_3^2)}~e^{-m_1^2/M^2}~e^{-m_2^2/M'^2}=\frac{1}{ 4 \pi^2}
\int_{s_{min}}^{s_0} ds \int_{u_{min}}^{u_0} du \:\rho_i^{OPE}(s,u,Q^2)
e^{-s/M^2}~e^{-u/M'^2},
\label{sumruleborel}
\eeq
From  (\ref{lagr}) it is clear that the above equation can be solved for the 
coupling 
$g_{M_1 M_2 M_3}$. Since the squared momenta of the mesons $M_1$ and $M_2$ 
($p^2$ and ${\pli}^2$  respectively) have been replaced by the Borel masses $M^2$ and 
${M'}^2$ and then fixed, the 
coupling $g$ will be a function only of the remaining Euclidean momentum $Q^2$, i.e., 
$g_{M_1 M_2 M_3}^{(M_3)} = g_{M_1 M_2 M_3}^{(M_3)} (Q^2)$ and this is what we call a 
form factor.  
The superscripts in parenthesis indicate the meson $M_3$ is off-shell.
At the point $Q^2 = - m_3^2$ the meson $M_3$ is on-shell and the form factor 
becomes the coupling constant. Of course one can not use $Q^2 = - m_3^2$ in
Eq.~(\ref{sumruleborel}) since the sum rule is only valid for $Q^2>0$. Therefore,
to obtain the coupling constant we will need to use some extrapolation procedure, 
that will be discussed in Sec.~\ref{extr}.

\subsection{Effective Lagrangians}
\label{effective}

Since the pioneering work of Matynian and M\"uller  \cite{mamu}, there has been an 
intense discussion concerning the details and properties of the effective 
Lagrangians which describe the interactions among charm mesons. Here we follow Refs. 
\cite{ko,osl}.
We write down the SU(4) chiral Lagrangian in terms of the mesonic fields:
\begin{equation}
{\cal L}_0=  \mbox{Tr} \left(\partial_{\mu}P^{\dagger} \partial^{\mu}{P}\right)
- \frac{1}{2}  \mbox{Tr} \left( F_{\mu\nu}^\dagger F^{\mu\nu}\right) ,
\end{equation}
where $F_{\mu\nu}= \partial_{\mu} V_{\nu} - \partial_{\nu}V_{\mu}$,
and $P $ and $V$ denote respectively the properly normalized $4 \times 4$ 
pseudoscalar and vector mesons matrices in SU(4) given by:
\begin{eqnarray}
P&=&\frac{1}{\sqrt{2}}\left(
\begin{array}{cccc}
\frac{\pi^0}{\sqrt{2}}+\frac{\eta}{\sqrt{6}}+\frac{\eta_c}{\sqrt{12}} & 
\pi^+ & K^+ &\bar{D^0} \\
\pi^- & -\frac{\pi^0}{\sqrt{2}}+\frac{\eta}{\sqrt{6}}+\frac{\eta_c}{\sqrt{12}} & 
K^0 & D^- \\
K^- & \bar{K^0} & -\sqrt{\frac{2}{3}}\eta+\frac{\eta_c}{\sqrt{12}} & D_s^- \\
D^0 & D^+ & D_s^+ &-\frac{3\eta_c}{\sqrt{12}} 
\label{P_matrix}
\end{array}
\right) , \\
V&=&\frac{1}{\sqrt{2}}\left(
\begin{array}{cccc}
\frac{\rho^0}{\sqrt{2}}+\frac{\omega'}{\sqrt{6}}+\frac{J/\psi}{\sqrt{12}} & \rho^+ & 
K^{*+} &\bar{D^{*0}} \\
\rho^- & -\frac{\rho^0}{\sqrt{2}}+\frac{\omega'}{\sqrt{6}}+\frac{J/\psi}{\sqrt{12}} & K^{*0} & D^{*- }\\
K^{*-} & \bar{K^{*0}} & -\sqrt{\frac{2}{3}}\omega'+\frac{J/\psi}{\sqrt{12}} & D_s^{*-} \\
D^{*0} & D^{*+} & D_s^{*+} &-\frac{3J/\psi}{\sqrt{12}} , 
\end{array}
\right).
\label{V_matrix}
\end{eqnarray}
In order to obtain the interaction Lagrangian we  introduce the following minimal substitutions:
\begin{eqnarray}
\partial_{\mu} P   &\rightarrow&  \partial_{\mu}P -\frac{i g}{2}\left[V_{\mu},P\right] \\
F_{\mu\nu} &\rightarrow& \partial_\mu V_{\nu}- \partial_ \nu  
V_ \mu -\frac{i g}{2} \left[V_{\mu},V_{\nu}\right] .
\end{eqnarray}
After using the Hermiticity of $P$ and $V$ the resulting Lagrangian reduces to:
\begin{eqnarray}
{\cal L }&=& {\cal L}_0 + ig \mbox{Tr} \left( \partial^{\mu}P \left[P,V_\mu\right]\right)-\frac{ g^2}{4} \mbox{Tr} 
\left(\left[P, V_\mu\right]^2\right)  \nonumber \\
&&+ ig \mbox{Tr} \left(\partial^\mu V^\nu \left[ V_{\mu},V_{\nu} \right]\right)+\frac{g^2}{8} 
\mbox{Tr}\left(\left[V_\mu,V_\nu\right]^2\right) .
\label{lagran1}
\end{eqnarray}
The above Lagrangian accounts for vertices of the form $PPV$ and $VVV$, where $P$ and $V$ denote pseudoscalar and 
vector mesons respectively. 
In order to include the vertices of the form $PVV$, it is necessary to use the following anomalous three-particle 
Lagrangian \cite{osl}:
\begin{equation}
{\cal L}_{int}^a=-\frac{g_a^2N_c}{16\pi^2F_\pi}
\epsilon^{\mu\nu\alpha\beta}\mbox{Tr}\left(\partial_\mu 
V_\nu\partial_\alpha V_\beta P\right). 
\label{lagran2}
\end{equation}
Equations  (\ref{lagran1}) and (\ref{lagran2}) are used to write the relevant meson 
effective Lagrangians. For example, in the $PPV$ ($VVV$) case, we choose two (none) mesons
in (\ref{P_matrix}) and one (three) meson in (\ref{V_matrix}) and set all the other 
elements of these matrices to zero. We then substitute the resulting 
Equations (\ref{P_matrix}) and (\ref{V_matrix}) into (\ref{lagran1}) and obtain the 
interaction Lagrangian for one specific  three meson interaction vertex. The 
coupling constants appearing in these Lagrangians are  charge specific (for example:  
$ g_{D^+ D^- J/\psi}$) and are  functions of the universal SU(4) couplings $g$ and $g_a$, 
being thus  connected to each other. 
They are  listed in Table \ref{tab_su4} and their connection 
with $g$ and $g_a$   is shown.  Using a compact notation,  the relevant Lagrangians are:
\begin{eqnarray}
{\cal L}_{\pi D^*D}&=&ig_{\pi D^*D}D_{\mu}^*( \bar{D} \partial^{\mu}{\pi} -  
\partial^{\mu}\bar{D} {\pi}) \label{dsdpi} , \\
{\cal L}_{\psi D^*D}&=&g_{\psi D^{*}D}\epsilon^{\mu \nu \alpha \beta}
\partial_{\mu}\psi_{\nu}\left(\partial_{\alpha}{D}^{*}_{\beta}\bar{D}+D\partial_{
\alpha}\bar{D^{*}}_{\beta}\right)
\label{psidsd} , \\
{\cal L}_{\psi D D}&=&ig_{\psi DD}\psi^\mu
\left ( \partial_\mu D \bar{D} - D \partial_\mu \bar{D} \right )
\label{psidd} , \\
{\cal L}_{\pi D^* D^*}&=&-g_{\pi D^*D^*}\epsilon^{\mu \nu \alpha \beta}
\partial_{\mu}D^*_{\nu}{\pi}\partial_{\alpha}\bar{D^*}_{\beta}
\label{dsdspi} , \\
{\cal L}_{\psi D^* D^*}&=& ig_{\psi D^*D^*}~
\left [ \psi^\mu \left ( \partial_\mu D^{* \nu} \bar {D^*_\nu}
- D^{* \nu} \partial_\mu \bar {D^*_\nu} \right )\right.\nonumber ,  \\
&&+\left ( \partial_\mu \psi_\nu D^{*\nu} -\psi_\nu
\partial_\mu D^*_\nu
\right )\bar {D^{* \mu}}
+\left. D^{* \mu} \left ( \psi^\nu \partial_\mu \bar {D^*_\nu} -
\partial_\mu \psi_\nu \bar {D^{*\nu}} \right ) \right ]
\label{psidsds}  , \\
{\cal L}_{\rho DD}&=&-ig_{\rho DD}~ \rho^\mu
\left ( \partial_\mu D \bar{D}  - D\partial_\mu \bar{D} \right )
\label{rhodd} , \\
{\cal L}_{\rho D^*D^*}&=& ig_{\rho D^*D^*}~
\left [ \rho^\mu \left ( \partial_\mu D^{* \nu} \bar {D^*_\nu}
- D^{* \nu} \partial_\mu \bar {D^*_\nu} \right )\right.\nonumber , \\
&&+\left ( \partial_\mu \rho_\nu D^{*\nu} -\rho_\nu
\partial_\mu D^*_\nu \right )\bar {D^{* \mu}}
+\left. D^{* \mu} \left ( \rho^\nu \partial_\mu \bar {D^*_\nu} -
\partial_\mu \rho_\nu \bar {D^{*\nu}} \right ) \right ] , 
\label{rhodsds}\\
{\cal L}_{\rho D^{*}D}&=&g_{\rho D^{*}D}\epsilon^{\mu \nu \alpha \beta}
\partial_{\mu}\rho_{\nu}\left(\partial_{\alpha}{D}^{*}_{\beta}\bar{D}+D\partial_{
\alpha}\bar{D^{*}}_{\beta}\right)
\label{rhodds} 
\end{eqnarray}
For future applications we present below the following Lagrangians containing quartic 
vertices:
\beqa
{\cal L}_{\psi D D \pi}&=&ig_{\psi D D \pi} \, \epsilon^{\mu \nu \alpha \beta}
\psi_{\mu}\partial_{\nu}D\partial_{\alpha}{\pi}\partial_{\beta}\bar{D}
\label{psiddpi} , \\
{\cal L}_{\psi D^* D \pi}&=&-g_{\psi D^* D \pi}
\psi^{\mu}\left(D\pi\bar{D^*}_{\mu}+D^*_{\mu}\pi\bar{D}\right)
\label{psidsdpi} , \\
{\cal L}_{\psi D^* D^* \pi}&=&ig_{\psi D^* D^* \pi} \, \epsilon^{\mu \nu \alpha \beta}
\psi_{\mu}{D^*}_{\nu}\partial_{\alpha}\pi\bar{D^*}_{\beta}+ih_{\psi D^* D^* \pi}
\, \epsilon^{\mu \nu \alpha \beta}\partial_\mu \psi_\nu D^{*}_{\alpha}\pi{D^*}_{\beta}  ,
\label{psidsdspi} 
\eeqa
where it is understood that, when doing practical calculations, charges must be specified in 
the above expressions. As it will be seen in the next section, after having calculated 
all specific coupling constants, we can, using some  appropriate convention, write them 
all in terms of generic coupling constants $g_{M_1 M_2 M_3}$.

From the above Lagrangians  we can derive the Feynman rule for the vertex 
(the  amplitude $ \langle M_1 M_2 M_3 \rangle $) which, as it can be seen,  
depends  on 
the relevant polarization vector, momenta and on the coupling $g_{M_1 M_2 M_3}$. 
The latter is the 
unknown, which will be determined by solving the sum rule. For example, for 
the $D^*D\pi$ vertex we get:
\beq
\langle D^*(p)|\pi(q)D(p-q)\rangle=g_{D^*D\pi}(q^2)q_\mu\varepsilon^\mu(p)\; ,
\eeq
where the momentum assignment is specified in the brackets and 
$\varepsilon^\mu$ is the polarization vector of the $D^*$.

\subsection{Couplings}

\begin{table}[htb]
\centering
\begin{tabular}{cc}
\hline
Coupling & Normal Vertices \\ \hline
$g/(4\sqrt{3})$ & $\omega D^0\bar{D^0},\omega D^+D^-,\omega D^{*0}\bar{D^{*0}},
\omega D^{*+}D^{*-}$\\
&$\eta D^{*0}\bar{D^{*0}},\eta  D^\pm D^{*\mp}$ \\ \hline
$g/4$ & $\pi^0D^0\bar{D^{*0}},\pi^0D^\pm D^{*\mp}$\\
&$\rho^0 D^0 \bar{D^0},\rho^0D^+D^-,\rho^0 D^{*0}\bar{D^{*0}},\rho^0D^{*+}D^{*-}$\\ \hline
$g/(2\sqrt{2})$ & $ \pi^+D^0D^{*-},\pi^-\bar{D^0} D^{*+},\pi^+D^-D^{*0},\pi^-D^+\bar{D^{*0}}$ 
\\
&$\rho^+D^0D^-,\rho^-\bar{D^0}D^+,\rho^+D^{*0}D^{*-},\rho^-\bar{D^{*0}} D^{*+}$ \\ \hline 
$g/\sqrt{6}$ & $\eta_cD^0\bar{D^{*0}},\eta_cD^\pm D^{*\mp}$\\
& $J/\psi D^0\bar{D^0},J/\psi D^+D^-,J/\psi D^{*0}\bar{D^{*0}},J/\psi D^{*\pm} D^{*\mp}$ 
\\ \hline
& Anomalous Vertices \\ \hline
$\alpha/(4\sqrt{6})$&$\eta_c\rho^0\rho^0,\eta_c\rho^+\rho^-,\eta_c\omega \omega$\\
&$J/\psi\pi^0\rho^0,J/\psi\pi^\pm\rho^\mp,J/\psi \eta\omega$\\ \hline
$\alpha/(4\sqrt{3})$&$\eta\omega \omega, \eta D^{*0}\bar{D^{*0}},\eta D^{*+}D^{*-},
\omega D^0\bar{D^{*0}}, \omega \bar{D^0}D^{*0},\omega D^\pm D^{*\mp}$ \\ \hline
$\alpha/(2\sqrt{6})$&$\eta_c D^{*0}\bar{D^{*0}},\eta_c D^{*+}D^{*-},\eta_c J/\psi J/\psi$ \\ 
&$J/\psi D^0\bar{D^{*0}},J/\psi\bar{D^0}D^{*0},J/\psi D^\pm D^{*\mp}$\\ \hline
$\alpha/4$&$\pi^0D^{*0}\bar{D^{*0}},\pi^0D^{*+}D^{*-},\rho^0D^0\bar{D^{*0}}, 
\rho^0\bar{D^0}D^{*0},\rho^0D^\pm D^{*\mp} $\\ \hline
$\alpha/(2\sqrt{2})$&$\rho^+D^0D^{*-},\rho^-\bar{D^0}D^{*+},\rho^+ D^- 
D^{*0}, \rho^- D^+ \bar{D^{*0}} $\\
\hline
\end{tabular}
\caption{Coupling constants for the charmed three-meson vertices within the SU(4) scheme.  
$\alpha=g_a^2N_c/(16\pi^2F_\pi)$.}
\label{tab_su4}
\end{table}
Using the formulas developed in  Section \ref{effective} we can write all the 
specific coupling constants in terms of the universal couplings $g$ and $g_a$.  
The SU(4) scheme is shown in Table \ref{tab_su4}, with the help of which we can relate all  
the coupling constants of charged states among themselves.  Therefore, using one known 
coupling constant, such as, for example the experimentally measured 
$g_{ {D^*}^{+} D^{0} \pi^{-}}$, we can infer  all the  others. The couplings determined in 
this way are already enough to be used as input in the 
Lagrangians (\ref{dsdpi}) - (\ref{rhodds}), which 
are then used in  calculations of phenomenological interest.  
However, in the literature, one often finds a ``generic''  coupling constant, i.e. 
without  any charge specification, such as, for example $g_{ D^* D \pi}$. The choice  of the 
generic  $g_{ D^* D \pi}$  (among the the possible charged state couplings 
$g_{ {D^*}^{+} D^{0} \pi^{-}}$, $g_{ {D^*}^{0} D^{0} \pi^{0}}$,  ...etc) is a matter of 
convention, as mentioned explicitly in \cite{mamu}. For the purpose of comparison we define 
the generic coupling as follows. If there is an isovector meson in the vertex, the  generic 
coupling is the one with a neutral isovector. For example: 
$$
g_{ D^* D \pi} =  g_{ {D^*}^{\pm} D^{\mp} \pi^{0}} = g_{ {D^*}^{0} D^{0} \pi^{0}}
$$
From Table  \ref{tab_su4} the generic coupling can be obtained from the states with the 
charged isovector: 
$$
g_{ D^* D \pi} = \frac{1}{\sqrt{2}} \,  g_{ {D^*}^{-} D^{0} \pi^{+}} =  
\frac{1}{\sqrt{2}} \,  g_{ {D^*}^{+} D^{0} \pi^{-}}
$$
If there is no isovector in the vertex, all the states have the same coupling and there is 
no ambiguity. Then, for example:
$$
g_{ J/\psi D D } =  g_{ J/\psi D^0 D^0 } = g_{ J/\psi D^+ D^- }
$$
All the calculations discussed throughout this review were performed for  
charged isovector currents.

In the next section we shall discuss in more detail the criteria which  must be 
satisfied by the sum rules, in order to be reliable.

\section{Evaluation of the sum rules}
\label{sec_eval}

\subsection{Numerical inputs}

In the following subsections we will present numerical results.
In the quantitative aspect, QCDSR is not  like a model which contains 
free parameters to be adjusted by fitting  experimental data. The inputs for numerical 
evaluations are the following: {\it i}) the vacuum matrix elements of composite 
operators involving quarks and gluons which appear in the operator product 
expansion (\ref{cope}). These numbers, known as  condensates, contain all the 
non-perturbative component of the approach. They could,  in principle, be 
calculated in lattice QCD. In  practice they are estimated 
phenomenologically.  They are universal and, once 
adjusted to fit, for example,  the mass of a particle, they must have always 
that same value. They are the analogue for spectroscopy of the parton 
distribution functions in deep inelastic scattering \cite{texto}; {\it ii}) quark masses, 
which are extracted from many  different phenomenological analyses; 
{\it iii})  the threshold parameter $s_0$ is the energy  (squared) 
which characterizes the beginning of the continuum, as shown in (\ref{ansacom}). Typically the quantity 
$\sqrt{s_0} - m$  (where $m$ is the mass of the ground state particle) is the  
energy needed to excite the  particle to its first excited state with the same 
quantum  numbers. This number is not well known, but should lie between $0.3$ and
$0.8$ GeV. If larger deviations from this interval are needed, the calculation 
becomes suspicious.  All in all, in QCDSR  we do not have much freedom for choosing  numbers. 
In the calculations discussed below we will use the numerical inputs shown in Table  \ref{table1}.  
The masses and decay constants are taken from the literature \cite{PDG}. They might also be calculated with
QCDSR as discussed in  previous subsections. 
\begin{table}[htb]
\centering
\begin{tabular}{cc}
\hline
Parameter & Value \\
\hline
$m_c (\GeV)$ & 1.27 $\pm$ 0.1 \\
$m_{D} (\GeV)$ &  1.86  \\
$m_{D^*} (\GeV) $   &  $2.01 $ \\
$m_{\rho} (\GeV)$ &  0.775  \\
$m_\psi   (\GeV) $  & $  3.1 $ \\
$f_{J/\psi}     (\GeV)$ &  $  0.405 \pm 0.015 $ \\
$f_{D^*} (\GeV)$ &  0.24 $\pm$ 0.02  \\
$f_{D} (\GeV)$ & 0.18 $\pm$ 0.02 \\
$f_{\rho} (\GeV)$ &  0.160 $\pm$ 0.005    \\
$f_{\pi} (\GeV)$ & 0.131  $\pm$  0.001 \\
$\langle \bar q q \rangle (\GeV)^3$  &  $(-0.23 \pm 0.01)^3$ \\
$\langle g^2G^2 \rangle   (\GeV)^4 $   &  $ 0.88 \pm 0.3 $  \\
\hline
\end{tabular}
\caption{Parameters used in the calculation with their errors.}
\label{table1}
\end{table}

\subsection{Borel stability}

As it was  mentioned above we perform two Borel transforms introducing the two Borel 
masses  $M^2$  and  ${M'}^2$.  The sum rules, expressing 
the interesting quantities as a function of the Borel parameters,  must be as much 
independent of these parameters as possible. An extensive check of this dependence has 
been carried out in \cite{00,01,02,02r,05,05r,05i,08,10}. For the sake of illustration 
we present below some 
numerical analysis taking the vertex $J/\psi D^* D$ as working example. 
For the continuum thresholds we have used $s_0=(m_\psi+\Delta_s)^2$ 
and $u_0=(m_{D(D^*)}+\Delta_u)^2$ for the sum rule where $D(D^*)$ is 
off-shell, and $s_0=(m_{D^*}+\Delta_s)^2$ 
and $u_0=(m_{D}+\Delta_u)^2$ for the sum rule where $J/\psi$ is off-shell.
We take $\Delta_s=\Delta_u=(0.5\pm0.1)\GeV$. 
\begin{figure}[htb]
\begin{center}
\epsfxsize=9cm
\leavevmode
\hbox{\epsffile{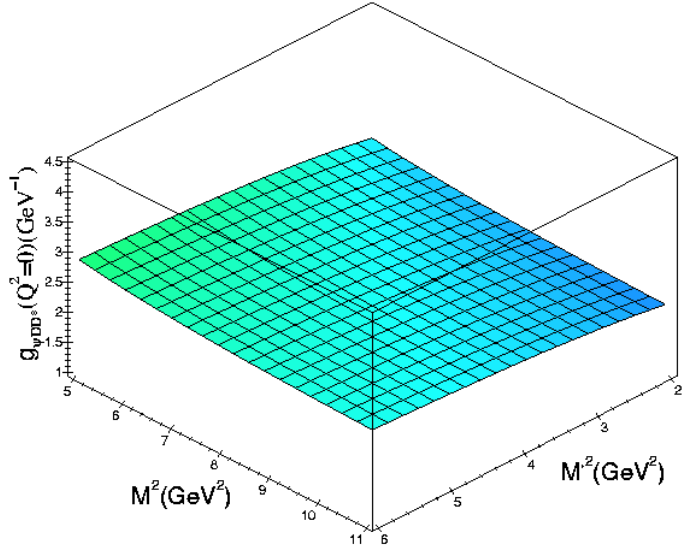}}
\end{center}
\protect\caption{$M^2$ and $\mli$ dependence of 
$g_{\psi DD^*}^{(D^*)}(Q^2=0)$.}
\label{f_psidds1}
\end{figure}
We first discuss the $J/\psi DD^*$ form factor with an off-shell $D^*$ 
meson.  Fixing $Q^2=-q^2=0$ and $\Delta_s=\Delta_u=0.5~\GeV$
we show in Fig.  \ref{f_psidds1} the Borel dependence of the form 
factor
$g_{\psi DD^*}^{(D^*)}(0)$. We see that we get a very good stability for 
the form factor as a function of the two independent Borel parameters in the 
considered Borel regions. 

\begin{figure}[h]
\begin{center}
\epsfxsize=9cm
\leavevmode
\hbox{\epsffile{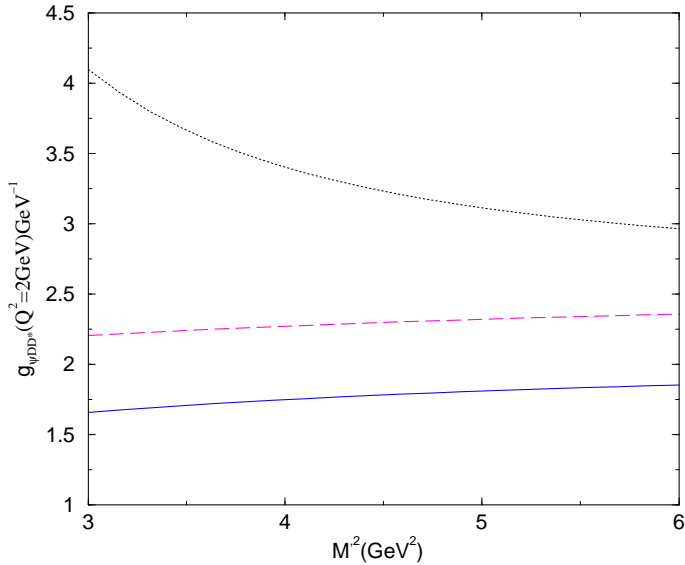}}
\end{center}
\protect\caption{$\mli$ dependence of $g_{J/\psi DD^*}^{(D^*)}$ (solid 
line), $g_{J/\psi DD^*}^{(D)}$ (dashed line) and  $g_{J/\psi DD^*}^{(J/
\psi)}$ 
(dotted line) for $Q^2=2.0\,\GeV^2$.}
\label{f_psidds3.eps}
\end{figure}

Since the dependence of the form factor on $M^2$  and  ${M'}^2$ is weak, 
there is no need to continue our analysis with these two independent variables. 
We can relate one to another and use only one variable. One commonly used relation is:
\beq
\frac{\mli}{M^2} = \frac{m_{M_1}^2}{ m_{M_2}^2} ,
\label{m2ml2}
\eeq
where  $M_1$ and $M_2$ are the mesons with $P^2=-p^2\rightarrow M^2$ and 
$P'^2=-{\pli}^2\rightarrow M'^2$ respectively. If the meson $M_1$  is light and 
the meson $M_2$ is heavy, the above equation is replaced by: 
\beq
\frac{\mli}{M^2} =  \frac{m_{M_1}^2}{(m_{M_2}^2 - m^2_c)} .
\label{m2ml2m}
\eeq
Other relations between  $M^2$  and  ${\mli}^2$ are possible, such as:
\beq
{\mli} =  a M^2 +  b  .
\label{baip}
\eeq
In \cite{baip} a comparative  analysis was performed leading to the conclusion that 
all these  forms  are acceptable. In what follows we will discuss results obtained with 
(\ref{m2ml2}) and with (\ref{m2ml2m}). We show, in Fig.~\ref{f_psidds3.eps}, the behavior of 
the form factors $g_{J/\psi DD^*}(Q^2)$ at $Q^2=2~\GeV^2$ 
as a function of the Borel mass $\mli$. 
The solid line gives 
$g_{J/\psi DD^*}^{(D^*)}$ at a fixed ratio $\mli/M^2={m_{D}^2
/m_\psi^2}$. The dashed line gives
$g_{J/\psi DD^*}^{(D)}$ at a fixed ratio $\mli/M^2={m_{D^*}^2/
m_\psi^2}$, and the dotted line gives $g_{J/\psi DD^*}^{(J/\psi)}$ 
at a fixed ratio $\mli/M^2={m_{D}^2/ m_{D^*}^2}$.
We can see that the QCDSR results for $g_{J/\psi D D^*}^{(D^*)}$ and
$g_{J/\psi D D^*}^{(D)}$ are very stable
in the interval $3 \leq  \mli  \leq 6 \,\GeV^2$. In the case of 
$g_{J/\psi D D^*}^{(J/\psi)}$ the stability is not as good as for the other
form factors, but it is still acceptable.

\subsection{OPE convergence}

As mentioned in the previous sections the operator product expansion should always be 
convergent. However sometimes this convergence is not so fast. In this section we 
present some examples, encountered in our calculations, which illustrate extreme cases 
ranging from a very fast convergence, where the second term is already negligible, to 
a slower convergence. Fortunately, in these latter cases, the obtention of a very weak 
dependence of the form factor on the Borel mass (i.e., the ocurrence of a stability 
``plateau'') indicates that the sum rule  is still enough convergent.  

\begin{figure}[h]
\begin{center}
\epsfxsize=9cm
\leavevmode
\hbox{\epsffile{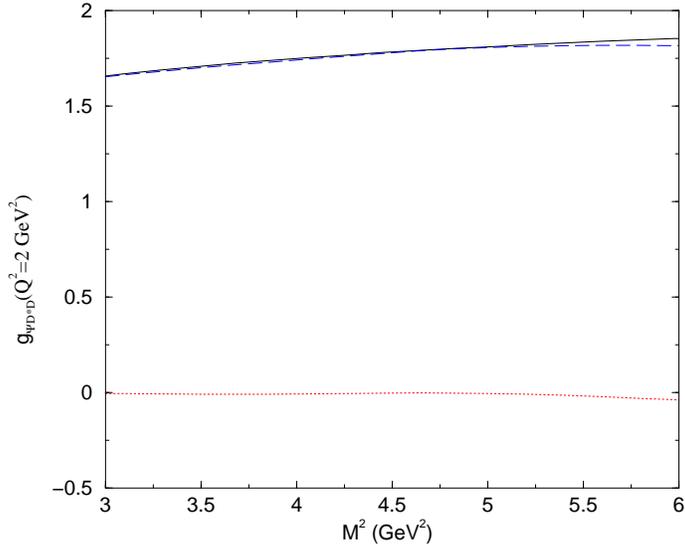}}
\end{center}
\protect\caption{$\mli$ dependence of perturbative contribution (dashed line)
and the gluon condensate contribution (dotted line) to 
$g_{J/\psi DD^*}^{(D^*)}$ at $Q^2=2.0\,\GeV^2$. The solid line gives the
final result for the form factor.}
\label{f_psidds2}
\end{figure}

\begin{figure}[h]
\begin{center}
\epsfxsize=9cm
\leavevmode
\hbox{\epsffile{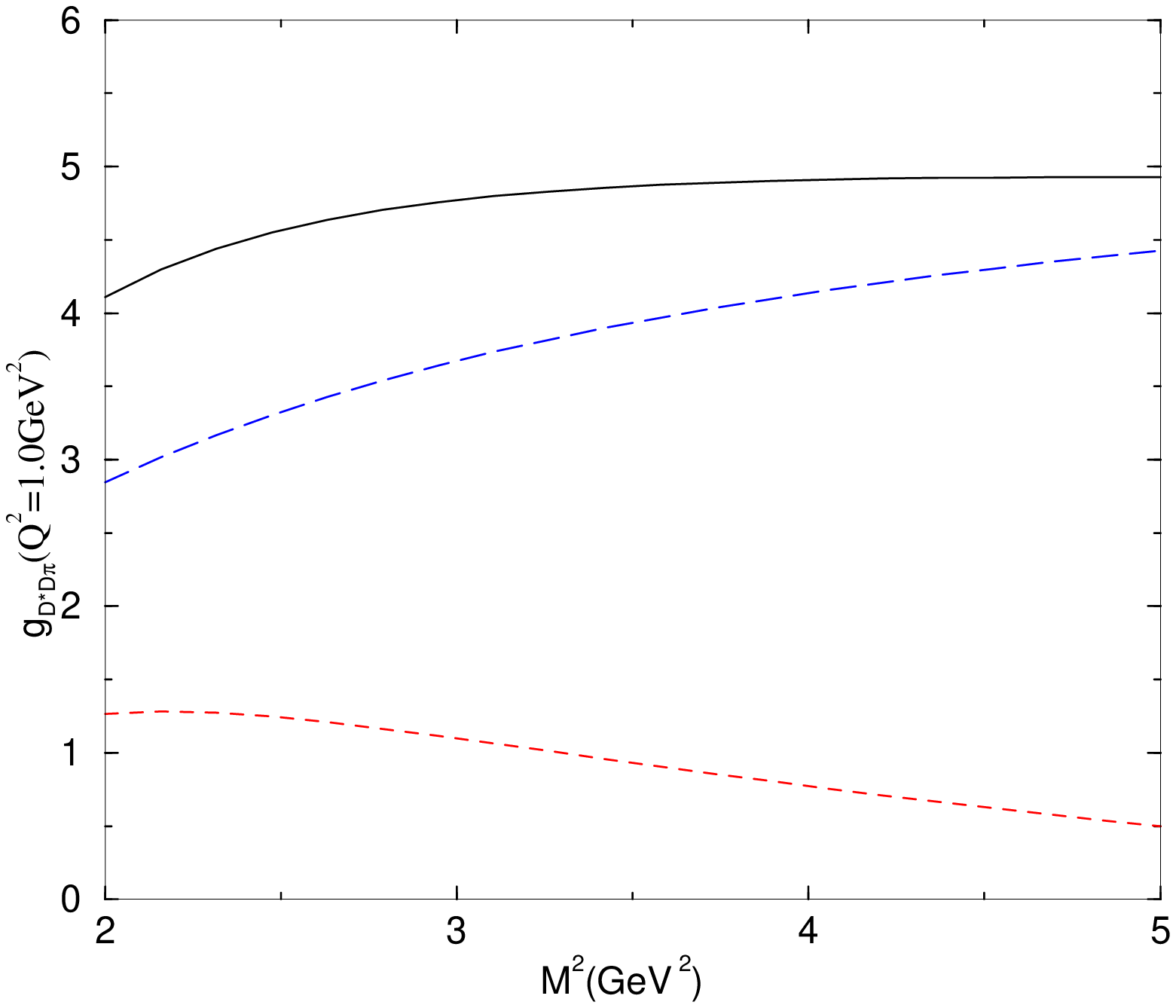}}
\end{center}
\protect\caption{$D^* D \pi$ form factor as a function of the Borel mass $M^2$ for an 
off-shell pion.
Dotted, dashed and solid lines represent the contribution of the gluon condensate, of the 
perturbative term and their sum (total) respectively.}
\label{ope_dsdpi}
\end{figure}

\begin{figure}[h]
\begin{center}
\epsfxsize=10cm
\leavevmode
\hbox{\epsffile{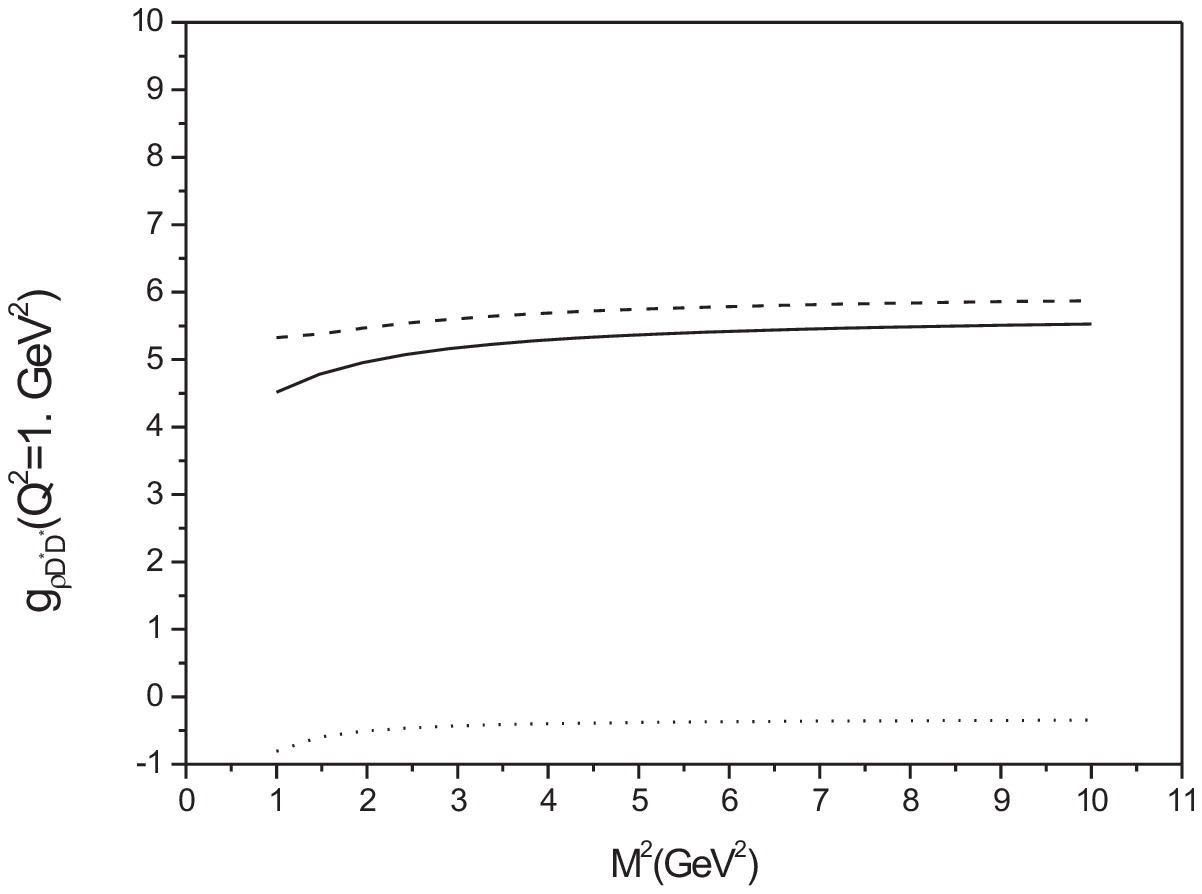}}
\end{center}
\protect\caption{$D^* D^* \rho$ form factor as a function of the Borel mass $M^2$ 
for an off-shell $D^*$. Dotted, dashed  and solid lines represent the 
contribution of the quark condensate, of the perturbative term and their 
sum (total) respectively.}
\label{ope_rhodsds}
\end{figure}

\begin{figure}[h]
\begin{center}
\epsfxsize=9cm
\leavevmode
\hbox{\epsffile{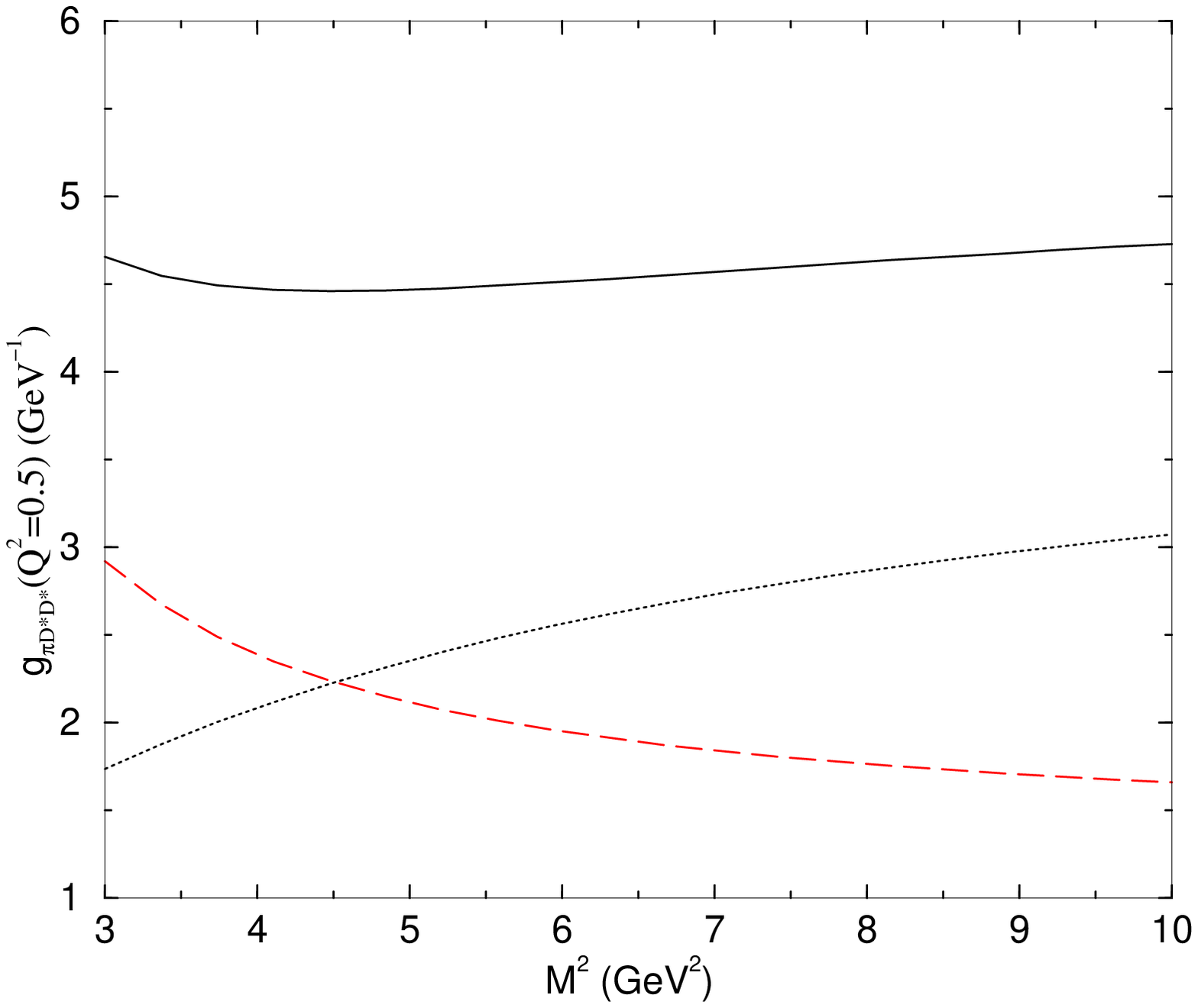}}
\end{center}
\protect\caption{$D^* D^* \pi$ form factor as a function of the Borel mass $M^2$ 
for an off-shell $D^*$. Dashed, dotted  and solid lines represent the 
contribution of the quark condensate, of the perturbative term and their 
sum (total) respectively.}
\label{ope_pidsds}
\end{figure}

In Fig.~\ref{f_psidds2} we show the perturbative (dashed line) and the gluon condensate
(dotted line) contributions to the form factor $g_{J/\psi DD^*}^{(D^*)}(Q^2)$ 
at $Q^2=2~\GeV^2$ as a function of the Borel mass $\mli$ at a fixed ratio 
$\mli/M^2={m_{D}^2/m_\psi^2}$.
We see that the gluon condensate contribution is negligible, when compared
with the perturbative contribution. 
The same kind of stability is obtained for 
other values of $Q^2$ and for the other two form factors \cite{05i,bjp}.
In Fig. \ref{ope_dsdpi} we show another example of OPE behavior, now in the 
$D^* D \pi$ vertex.  In this Figure we study the $D^* D \pi$ vertex, choosing the pion 
to be off-shell with $Q^2 = 1 $ GeV$^2$ and setting  $M'^2=M^2$. The solid line shows the 
sum of the first two relevant terms of the OPE, the perturbative term and the gluon 
condensate. It is reassuring to observe that the former is much larger than the latter 
and also that their sum is very flat over a wide range of values of $M^2$. 
Although here the dominance of the perturbative term over the gluon condensate is less 
pronounced, we observe an interesting feature, namely, that the inclusion of the gluon 
condensate  contributes to the Borel stability of the sum rule \cite{00,02r}.
In Figs.  \ref{ope_rhodsds} and   \ref{ope_pidsds}  
we show a comparison between the perturbative term and quark 
condensate term. They illustrate how small (Fig.  \ref{ope_rhodsds})
and how large (Fig.    \ref{ope_pidsds}) the effects of the quark condensate  
can be. 

\subsection{Pole versus continuum}

\begin{figure}[ht]
\begin{center}
\epsfxsize=9cm
\leavevmode
\hbox{\epsffile{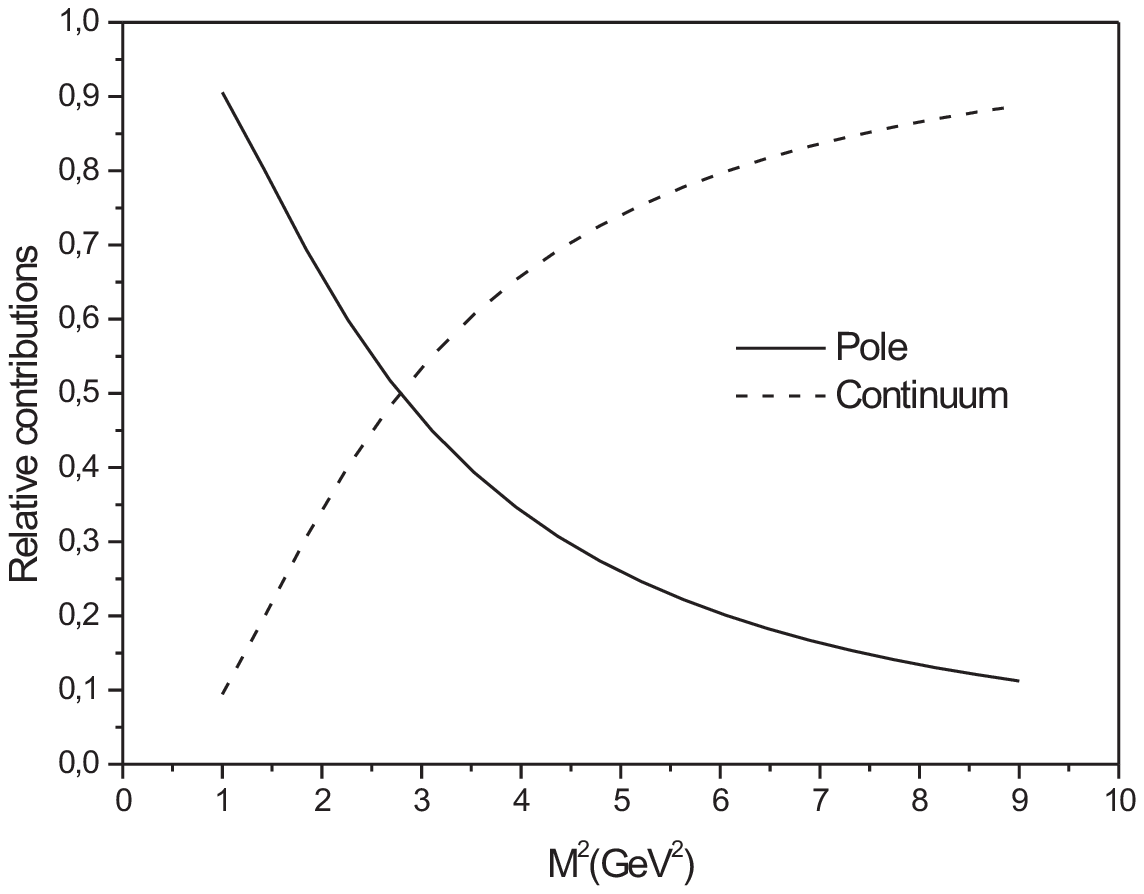}}
\end{center}
\protect\caption{Pole (solid line) and continuum (dashed line) contribuition to  
$g^{(\rho)}_{\rho D^*D^*}(Q^2=1\GeV^2)$, 
as a function of the Borel mass $M^2$.}
\label{pole_cont_rho}
\end{figure}

\begin{figure}[ht]
\begin{center}
\epsfxsize=9cm
\leavevmode
\hbox{\epsffile{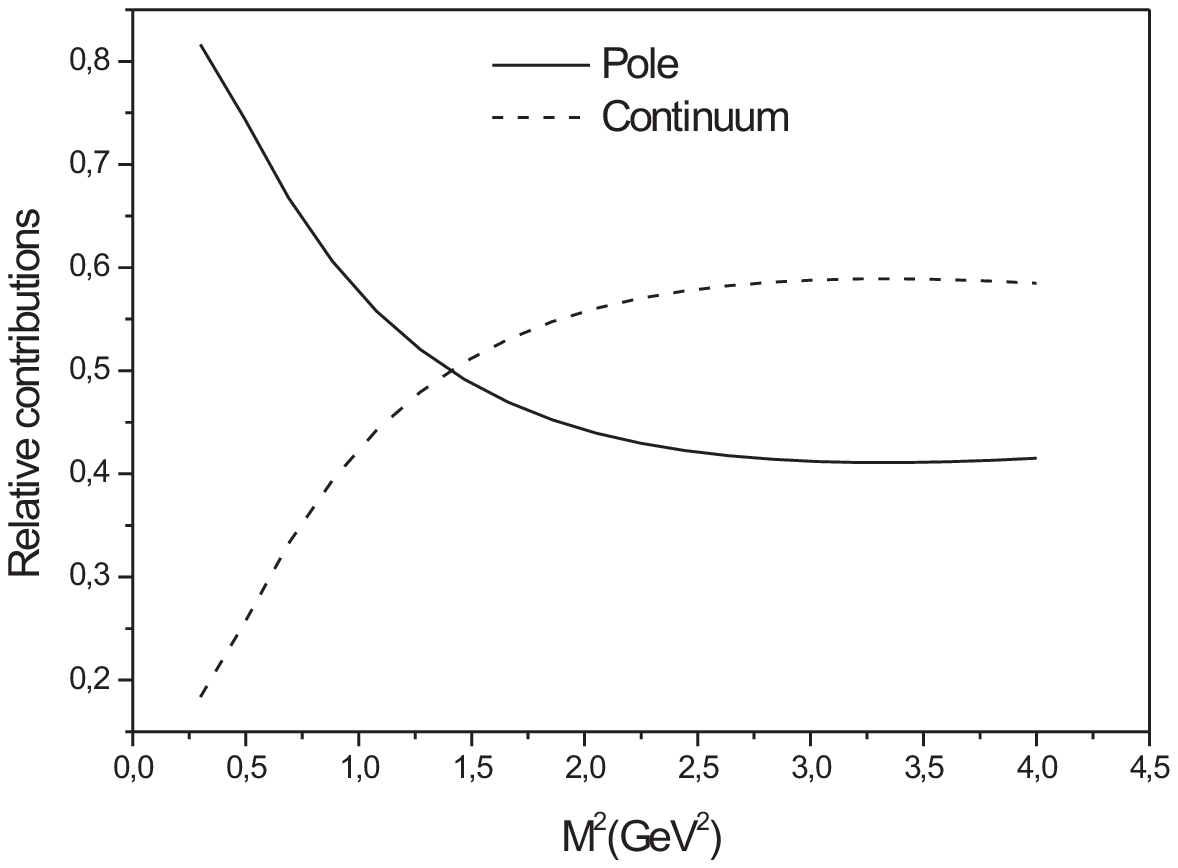}}
\end{center}
\protect\caption{Pole versus continuum contributions  to  $g^{(D^*)}_{\rho D^*D^*}(Q^2=1\GeV^2)$  
as a function of the Borel mass $M^2$.}
\label{pole_cont_d}
\end{figure}

In the study of the two-point function a crucial assumption is the form of the 
spectral density, Eq. (\ref{den}). Moreover, it is also crucial that the dispersion 
integral be dominated by the pole contribution and not by the continuum. In the case
of the two-point function, if this condition is not satisfied we can not compute the 
mass of the particle. A similar condition must hold for the three-point function. If
it is not dominated by the pole contribution, this could be an indication that there 
is more than one off-shell particle  in the vertex. 
In the case of the two-point function the pole dominance can be tested with the
help of (\ref{disp}).  The pole and continuum contributions are  given by:
\beq
\Pi(q^2)_{pole}= -  \int_{s_{min}}^{s_0} ds\, {\rho^{OPE}(s)\over q^2-s+i\epsilon} ,
\label{dispole}
\enq
\beq
\Pi(q^2)_{cont}= -  \int_{s_0}^{\infty} ds\, {\rho^{OPE}(s)\over q^2-s+i\epsilon} .
\label{discont}
\enq
We can perform a Borel transform in the above expressions and plot the relative 
contributions, which are given by:
\beq
\mbox{Pole} = \frac{\Pi(M^2)_{pole}}{\Pi(M^2)_{pole} + \Pi(M^2)_{cont}} , 
\label{ratpole}
\enq
\beq
\mbox{Continuum} = \frac{\Pi(M^2)_{cont}}{\Pi(M^2)_{pole} + \Pi(M^2)_{cont}}  .
\label{ratcont}
\enq
A similar analysis can be carried out for the three-point function.  
Starting with 
(\ref{dis}) and choosing one of the $i$ tensor structures we can write:
\begin{equation}
F^{OPE}_{pole} =-\frac{1}{4 \pi^2}\int_{s_{min}}^{s_0} ds
\int_{u_{min}}^{u_0} du \:\frac{\rho^{OPE}(s,u,Q^2)}{(s-p^2)(u-{\pli}^2)} ,
\label{three_pole}
\end{equation}
\begin{equation}
F^{OPE}_{cont} =-\frac{1}{ 4 \pi^2} \int_{s_0}^{\infty} ds
\int_{u_0}^{\infty} du \:\frac{\rho^{OPE}(s,u,Q^2)}{(s-p^2)(u-{\pli}^2)} ,
\label{three_cont}
\end{equation}
and then, performing  a double Borel transform  and using (\ref{m2ml2}) or (\ref{m2ml2m}),   
we write the two relative contributions as:
\beq
\mbox{Pole} = \frac{F^{OPE}(M^2)_{pole}}{F^{OPE}(M^2)_{pole} + F^{OPE}(M^2)_{cont}} , 
\label{rathree_pole}
\enq
\beq
\mbox{Continuum} = \frac{F^{OPE}(M^2)_{cont}}{F^{OPE}(M^2)_{pole} + F^{OPE}(M^2)_{cont}} . 
\label{rathree_cont}
\enq
In Figs. \ref{pole_cont_rho} and  \ref{pole_cont_d}, we show the pole-continuum analysis 
for the $D^* D^* \rho$ vertex \cite{08}. In Fig. \ref{pole_cont_rho} the off-shell particle is 
the $\rho$, with virtuality $Q^2 =1 $ GeV$^2$. In Fig. \ref{pole_cont_d}  the off-shell 
particle is the $D^{*}$ with $Q^2 =1 $ GeV$^2$. As it can be seen, for masses higher than 
$\simeq 2.5$ GeV$^2$ the sum rule is dominated by the continuum. This condition sets an  
upper limit for the Borel mass. A lower limit comes from imposing the OPE convergence, as 
it can be inferred from Fig. \ref{ope_dsdpi}.    This interval is called the ``Borel window'' 
and it does not always exist. Changing the value of $Q^2$ may help or hinder the existence 
of the Borel window. In the next sections we will present QCDSR results for the form 
factors. We will show curves of $g(Q^2)$ as a function of $Q^2$ for a fixed $M^2$.  
The points where we have results are restricted to a certain region (sometimes 
relatively narrow) of the $Q^2$ axis. One of the reasons for this limited range of 
applicability is the requirement of the pole dominance.  
\subsection{Continuum threshold effects}
 
\begin{figure}[htb]
\begin{center}
\epsfxsize=9cm
\leavevmode
\hbox{\epsffile{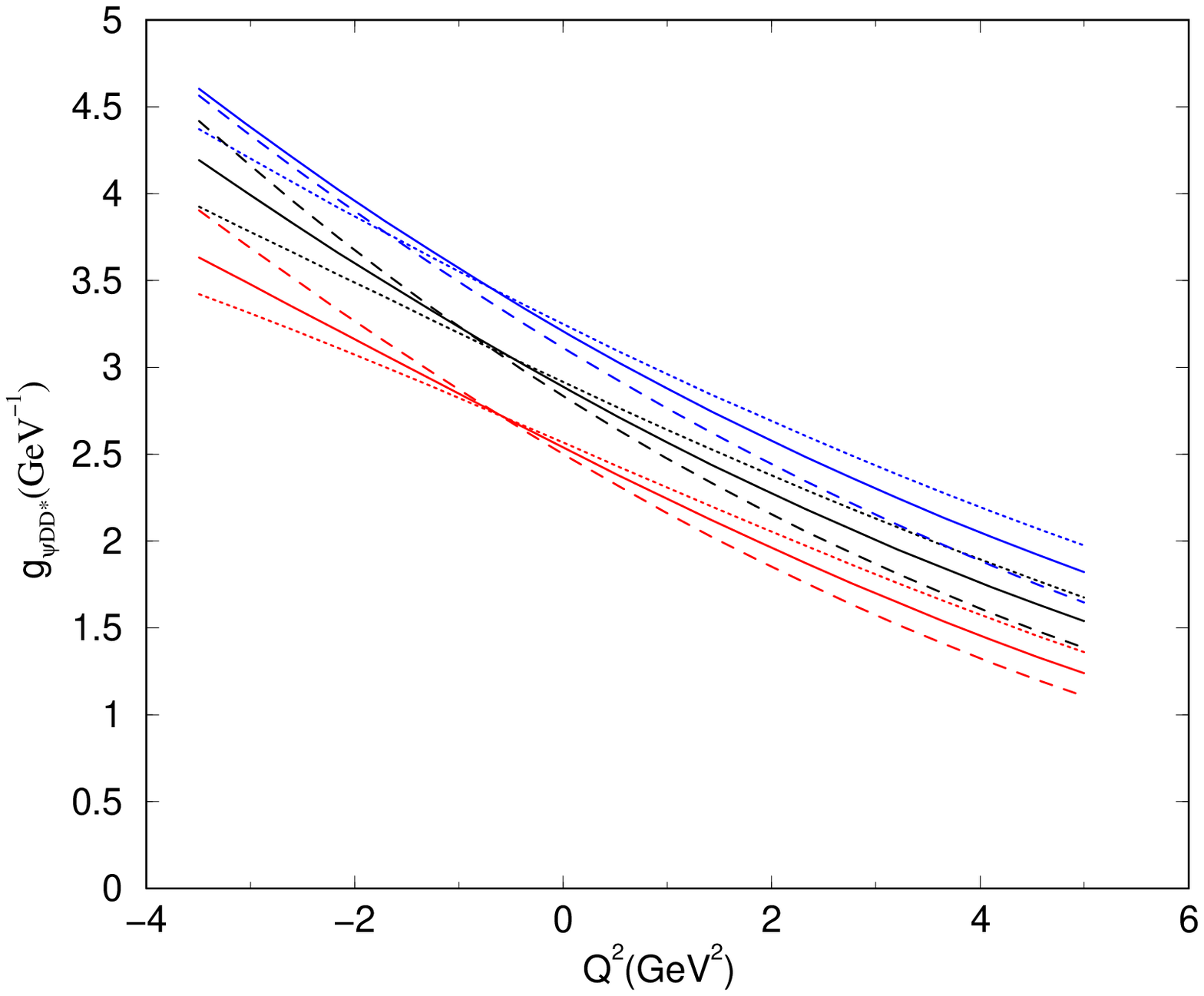}}
\end{center}
\protect\caption{Continuum threshold dependence of the form factor
$g^{(D)}_{\psi DD^{*}}(Q^2)$. The dashed solid and dotted lines give the 
parametrization of the QCDSR results for  $\Delta_s=0.4~\GeV,~0.5~\GeV$
and $0.6~\GeV$ respectively. The lower, intermediate and upper set of curves 
show the results for $\Delta_u=0.4~\GeV,~0.5~\GeV$
and $0.6~\GeV$ respectively.}
\label{f_psidds5.eps}
\end{figure}

\begin{figure}[h]
\begin{center}
\epsfxsize=9cm
\leavevmode
\hbox{\epsffile{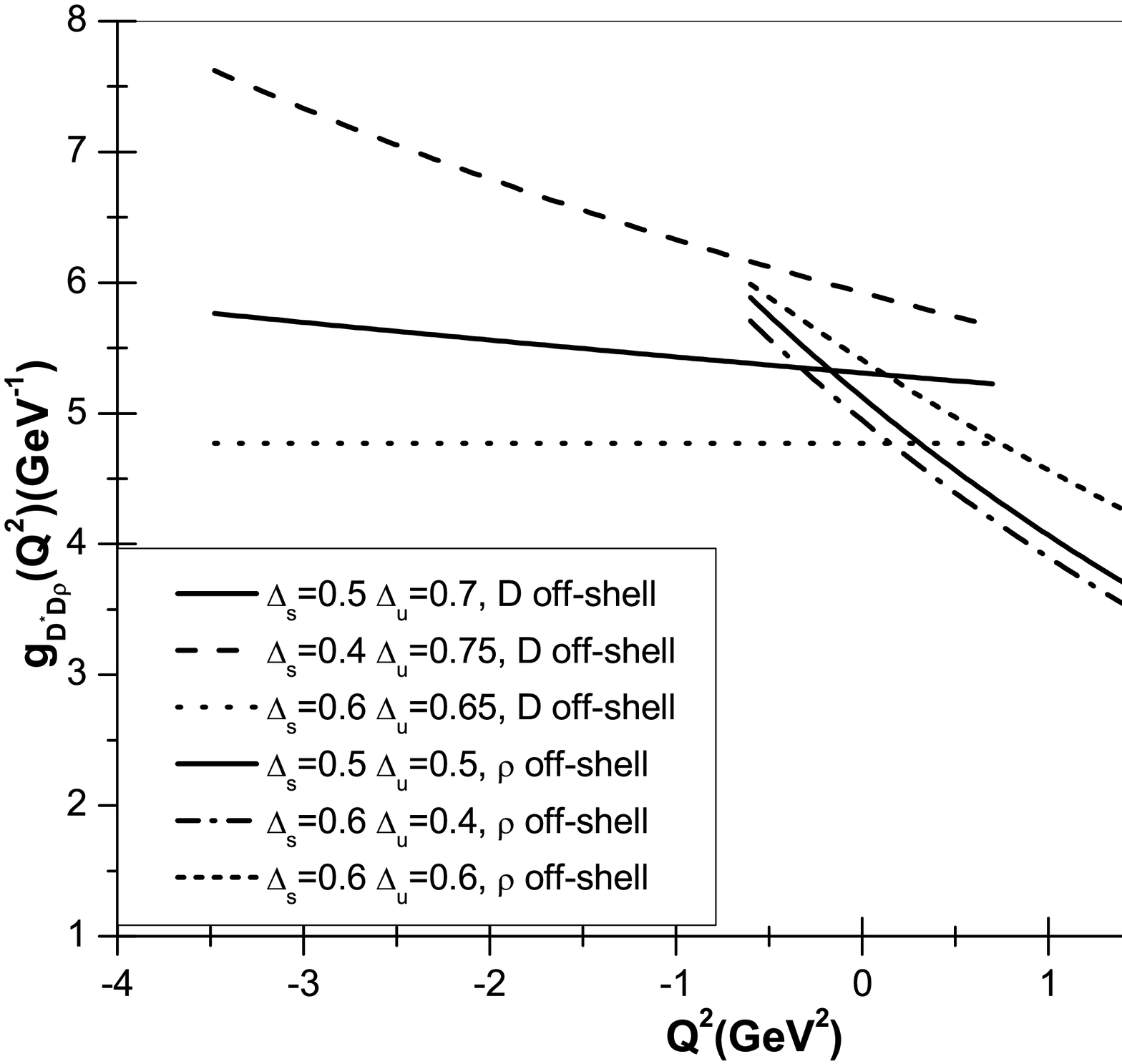}}
\end{center}
\protect\caption{Dependence of the $  D^* D \rho $ form factors on the continuum thresholds.
The steeper lines correspond to $\rho$ off-shell. The others are for D off-shell.}
\label{dsdrho_ex}
\end{figure}

The results showed in Figs. \ref{f_psidds3.eps},  \ref{f_psidds2}, 
\ref{ope_dsdpi}, \ref{pole_cont_rho} and \ref{pole_cont_d} were 
obtained using $\Delta_s=\Delta_u=0.5~\GeV$.
In Fig.~\ref{f_psidds5.eps} we use the form factor $g^{(D)}_{J/\psi DD^{*}}(Q^2)$ 
to illustrate the dependence of our results on the continuum thresholds. 
In the Figure there are three sets of three curves. They show fits of the QCD sum 
rule results. The lower, intermediate and upper sets were obtained using 
$\Delta_u=0.4~\GeV,~0.5~\GeV$ and $0.6~\GeV$ respectively. The dashed,  solid and 
dotted lines in each set were obtained using $\Delta_s=0.4~\GeV,~0.5~\GeV$
and $0.6~\GeV$ respectively. In this case we can see that the dispersion in the 
region $0\leq Q^2\leq4.5~\GeV^2$, where we have the QCDSR points, does not
lead to a bigger dispersion at $Q^2=-m_D^2$, where the coupling constant
is extracted. As it will be discussed in the next subsection,  the extrapolation procedure 
used here does not necessarily amplify the uncertainties. In fact, in some cases the 
uncertainties can be even damped  as we move to the time-like region. An example of this 
damping is shown in 
Fig. \ref{dsdrho_ex}, where we plot fits of the QCDSR results for the form factor of
the $D^* D \rho$ vertex. The two sets of lines correspond to a $\rho$ off-shell (steeper
lines) and to a $D$ off-shell. Each line corresponds to a different choice of the continuum  
threshold parameters.  As it can be seen, this is one of the major sources of uncertainties. 

\subsection{Choice of the structure}

As it was discussed in the previous sections,  the explicit evaluation of the correlation
functions both in the OPE and in the phenomenological side leads to expressions written 
in terms of several tensor structures. We can write a sum rule identifying the coefficients 
of each structure and hence we have as many sum rules as structures. In principle all the 
sum rules are equivalent and should yield the same final results. In practice however, 
the truncation of the OPE affects different structures (and the corresponding  sum rules) in
different ways. Consequently some structures lead to sum rules which are more stable. 
In the simplest cases, such as in the $D^* D \rho$ vertex, we have only one structure and,  
using the notation of (\ref{struc}) and (\ref{strucn}), the correlator  
(both in the phenomenological and the OPE descriptions) is written as:
\beq
\Gamma_{\mu\nu}(p,\pli)=F(p^2,{\pli}^2,q^2) \, 
\epsilon_{\alpha\beta \mu\nu}p^\alpha{\pli}^\beta ,
\label{dsdrho_phen}
\eeq
and an analogous expression for the OPE side. In the most complicated cases, 
as in the $\rho D^* D^*$ or $J/\psi D^* D^*$ vertices,  the number of structures is fourteen
and the correlators (both in the phenomenological and the OPE descriptions) have the 
following tensor decomposition:
\begin{eqnarray}
\Gamma_{\mu\nu\alpha}(p,\pli)&=&
    F_1(p^2 , \ql , q^2) g_{\mu \nu} p_{\alpha} 
  + F_2(p^2,\ql, q^2) g_{\mu \alpha} p_{\nu} 
  + F_3(p^2,\ql , q^2) g_{\nu \alpha} p_{\mu}   \nonumber \\ 
&&+ F_4(p^2,\ql ,q^2) g_{\mu \nu} \pli_{\alpha}  
  + F_5(p^2, \ql ,q^2) g_{\mu \alpha} \pli_{\nu} 
  + F_6(p^2,\ql ,q^2) g_{\nu\alpha} \pli_{\mu}   \nonumber \\ 
&&+ F_7(p^2,\ql ,q^2) p_{\mu} p_{\nu} p_{\alpha}
  + F_8(p^2,\ql ,q^2) \pli_{\mu} p_{\nu} p_{\alpha} + F_9(p^2,\ql ,q^2) p_{\mu} \pli_{\nu} 
  p_{\alpha} \nonumber \\ 
&&+ F_{10}(p^2,\ql ,q^2) p_{\mu} p_{\nu} \pli_{\alpha} 
  + F_{11}(p^2,\ql ,q^2) \pli_{\mu} \pli_{\nu} p_{\alpha}
  + F_{12}(p^2,\ql ,q^2) \pli_{\mu} p_{\nu} \pli_{\alpha}  \nonumber \\
&&+ F_{13}(p^2,\ql ,q^2) p_{\mu} \pli_{\nu} \pli_{\alpha} 
  + F_{14}(p^2,\ql ,q^2) \pli_{\mu} \pli_{\nu} \pli_{\alpha} .
  \label{trace}  
\end{eqnarray}
The above equation is the most general expression that can be written with three Lorentz
indices $\mu,~\nu$ and $\alpha$. It makes use of two independent four vectors,
$p$ and $q$, and the metric tensor. It contains fourteen Lorentz structures with fourteen 
invariant functions $F_i$. Although all the  Lorentz structures are independent, 
the fourteen invariant functions can not be independent due to current conservation. 
Enforcing current conservation reduces the number of independent invariant functions. 
Usually we may take advantage of current conservation and write sum rules which are simpler. 
The best example is the two-point function of the vector mesons $\rho$ and $J/\psi$ where, 
instead of studying separately the two possible independent structures $g_{\mu\nu}$ and $q_\mu q_\nu$, 
we may use  a combination of them which is manifestly conserved:
\begin{equation}
\Pi_{\mu\nu}(q)=\Pi(q^2)(-g_{\mu\nu}q^2+q_\mu q_\nu) ,
\label{pro}
\end{equation}
since $q^\mu\Pi_{\mu\nu}=0$. The sum rule is thus written for the scalar function
$\Pi(q^2)$. Inspite of technical advantages of using (\ref{pro}), we  may
still choose to work with one single structure, either $g_{\mu\nu}$ or 
$q_\mu q_\nu$. In the case of (\ref{trace}) current conservation yiels identities such as 
\begin{equation}
p^\mu\Gamma_{\nu\alpha\mu}(p, \pli)=0  ,
\label{e1}
\end{equation}
or
\begin{equation}
{\pli}^\alpha\Gamma_{\nu\alpha\mu}(p, \pli)=0 , 
\label{e2}
\end{equation}
depending on the momentum of the $J/\psi$ or $\rho$ meson.  
With the help of expressions (\ref{e1}) or (\ref{e2}) we might rewrite the sum rules in terms of 
combinations of different structures. However there is nothing wrong in working with  individual 
Lorentz structures, as done in \cite{05} and \cite{08}.

\begin{figure}[h]
\begin{center}
\epsfxsize=9cm
\leavevmode
\hbox{\epsffile{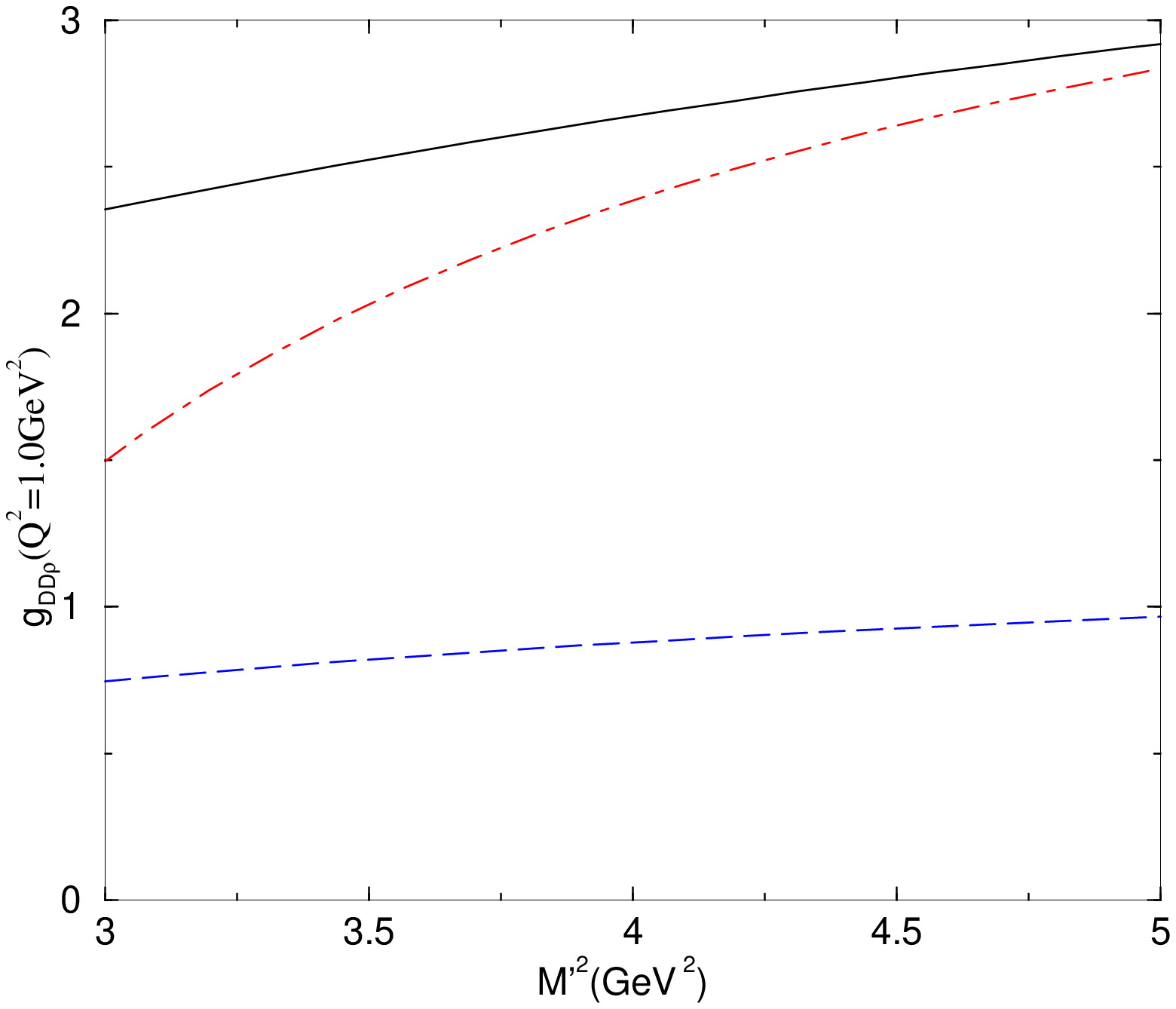}}
\end{center}
\protect\caption{$\mli$ dependence of the $DD\rho$ form factors at $Q^2=1\,\GeV^2$ 
for $\Delta_s=\Delta_u=0.5\,\GeV$. The dashed line gives the QCDSR result
for $g_{DD\rho}^{(\rho)}(Q^2)$ and the dot-dashed and solid lines give 
the QCDSR results for $g_{DD\rho}^{(D)}(Q^2)$ in the $p_\mu$ and $\pli_\mu$
structures respectively.}
\label{ddrho_struc}
\end{figure}

\begin{figure}[h]
\begin{center}
\epsfxsize=10cm
\leavevmode
\hbox{\epsffile{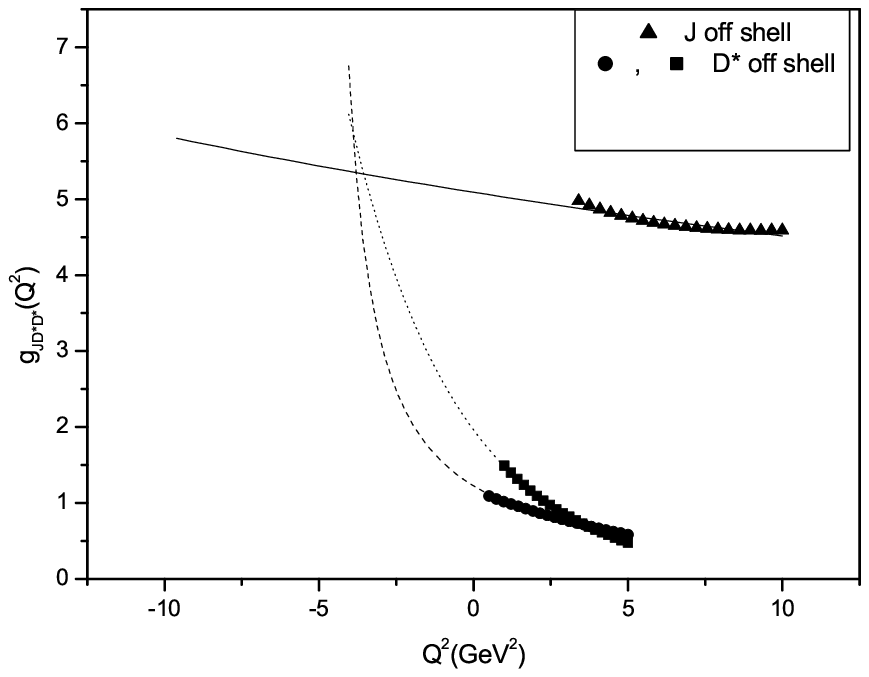}}
\end{center}
\protect\caption{$g^{(D^*)}_{J/\psi D^*D^*}$ (circles and squares) and 
$g^{(J/\psi)}_{J/\psi D^*D^*}$ (triangles) form factors as a function of
$Q^2$ from the QCDRS calculation of this work. The solid and dashed lines 
correspond to the  monopole parametrization of the QCDSR data, using 
two different structures, and the  dotted line corresponds to the exponential parametrization.}
\label{psidsds_struc}
\end{figure}

In Fig. \ref{ddrho_struc} we show the typical difference between results obtained with 
different structures. In the Figure we see the form factors of the $\rho D D$ vertex, 
where we have only two structures $p_{\mu}$ and $\pli_{\mu}$. The dashed line shows the 
case of an off-shell $\rho$ meson. In this case the two structures give the same result. 
The solid and dot-dashed lines 
refer to the case of  an off-shell D meson and they show the same quantity computed in the 
two different structures.  We can observe a clear difference between the two  lines, 
especially in the low Borel mass region. Therefore the choice of structures deserves 
attention, since it may be an additional source of uncertainties.    
In Fig. \ref{psidsds_struc} we show the form factors of the $J/\psi D^* D^*$ vertex for 
fixed $M^2$ as a function of $Q^2$.  The dashed and dotted lines show results obtained 
with different structures. 

\section{ Form factors}
\label{sec_form}

In the previous section we have presented the conditions which must be satisfied for a 
sum rule to 
be considered reliable.  In this section we discuss a difficulty inherent to the calculation 
of coupling 
constants with QCDSR. The solution of (\ref{sumrule}) is numerical and restricted to a 
singularity-free region in the $Q^2$ axis, usually located in the space-like region. 
Therefore, in order 
to reach the
pole position, $Q^2 = - m_{M_{3}}^2$, we must fit the solution, finding a function
$g_{M_1 M_2 M_3}^{(M_3)} (Q^2)$ which is then extrapolated to the pole, yielding the coupling
constant. In the following subsections we introduce an extrapolation procedure and discuss 
how to improve it.

\subsection{The extrapolation procedure}
\label{extr}

In order to minimize the uncertainties associated with the extrapolation 
procedure, for each vertex  we perform the calculation twice, putting first one meson and 
then another meson off-shell, obtaining two form factors 
$g^{(M_1)}_{M_1 M_2 M_3} (Q^2)$ and $g^{(M_2)}_{M_1 M_2 M_3} (Q^2)$ and 
requiring that these  two functions have the same value at the respective poles. 
The superscripts in parenthesis indicate which meson is off-shell.

\begin{figure}
\begin{center}
\epsfxsize=9cm
\leavevmode
\hbox{\epsffile{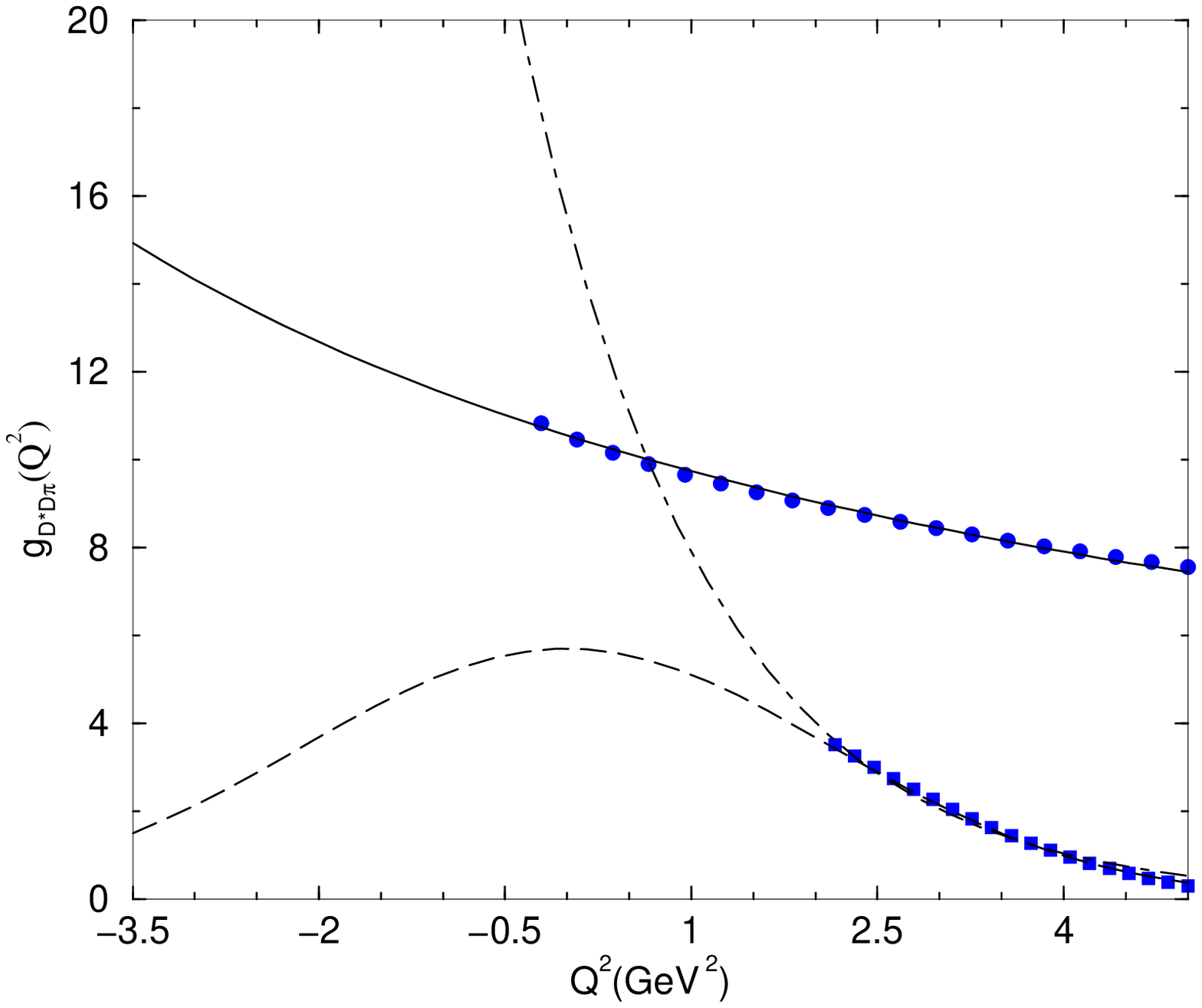}}
\end{center}
\protect\caption{$D^* D \pi$ form factor as a function of  $Q^2$ for an off-shell pion 
(dashed and dot-dashed lines) and an off-shell $D$ (solid line). Circles and squares are the 
results of numerical calculations and the curves are their fits.}
\label{fig2}
\end{figure}

In order to yield reliable results the sum rule (\ref{sumrule}) must satisfy the quality 
criteria discussed in the previous  subsections.  In first place (\ref{sumrule}) is a 
function of two Borel  masses $M^2$ and $M'^2$. A good sum rule is independent of the choice 
of these masses (it is ``Borel stable''),  showing a plateau when plotted as a 
function of $M^2$ or $M'^2$. Moreover, the OPE side is a series which must be convergent. 
We  choose a value for  $M^2$ and plot (\ref{sumrule}) as a function of $Q^2$, as it is  
shown in  Fig. \ref{fig2}. The squares and circles show the result of the numerical 
evaluation of the form factor $g(Q^2)$ as a function of $Q^2$ for the $D^* D \pi$ vertex with 
a pion (squares) and a $D$ (circles) off-shell. As it can be seen, at a certain (low) value of 
$Q^2$ the calculation stops, because at this point the stability and convergence criteria are 
no longer satisfied. From here on we have to extrapolate.  In  Fig. \ref{fig2} we show 
fits of the QCDSR results represented by the  lines. As it is illustrated in the case of an  
off-shell pion, the numerical points obtained with QCDSR can be fitted, 
with similar accuracy, by several forms 
which, when extrapolated to $Q^2=-m^2_{\pi}$,  will lead to very different points! 
This can be clearly seen by comparing the dashed and dash-dotted lines in  Fig.  
\ref{fig2}.  In order to reduce the freedom in the extrapolation and constrain the 
form factor we calculate and fit simultaneously the values of $g(Q^2)$ of the same 
$D^* D \pi$ vertex with the $D$ off-shell.
The results are shown in Fig. \ref{fig2} with circles fitted by the solid line. 
We perform the fits of the two sets of points (circles and squares) imposing the 
condition that 
{\it the two resulting parametrizations, when extrapolated to $Q^2=-m^2_{\pi}$ and  
$Q^2=-m^2_{D}$ go to the same value of  $g_{D^* D \pi}(Q^2)$}. This procedure is 
enough to reduce the uncertainties and, imposing this requirement leads to 
$g_{D^* D \pi} = 14.0 \pm 1.5$, which is consistent with the experimental value 
$g_{D^* D \pi} = 17.9 \pm 0.3 \pm 1.9$.

\begin{figure}[htb]
\begin{center}
\epsfxsize=9cm
\leavevmode
\hbox{\epsffile{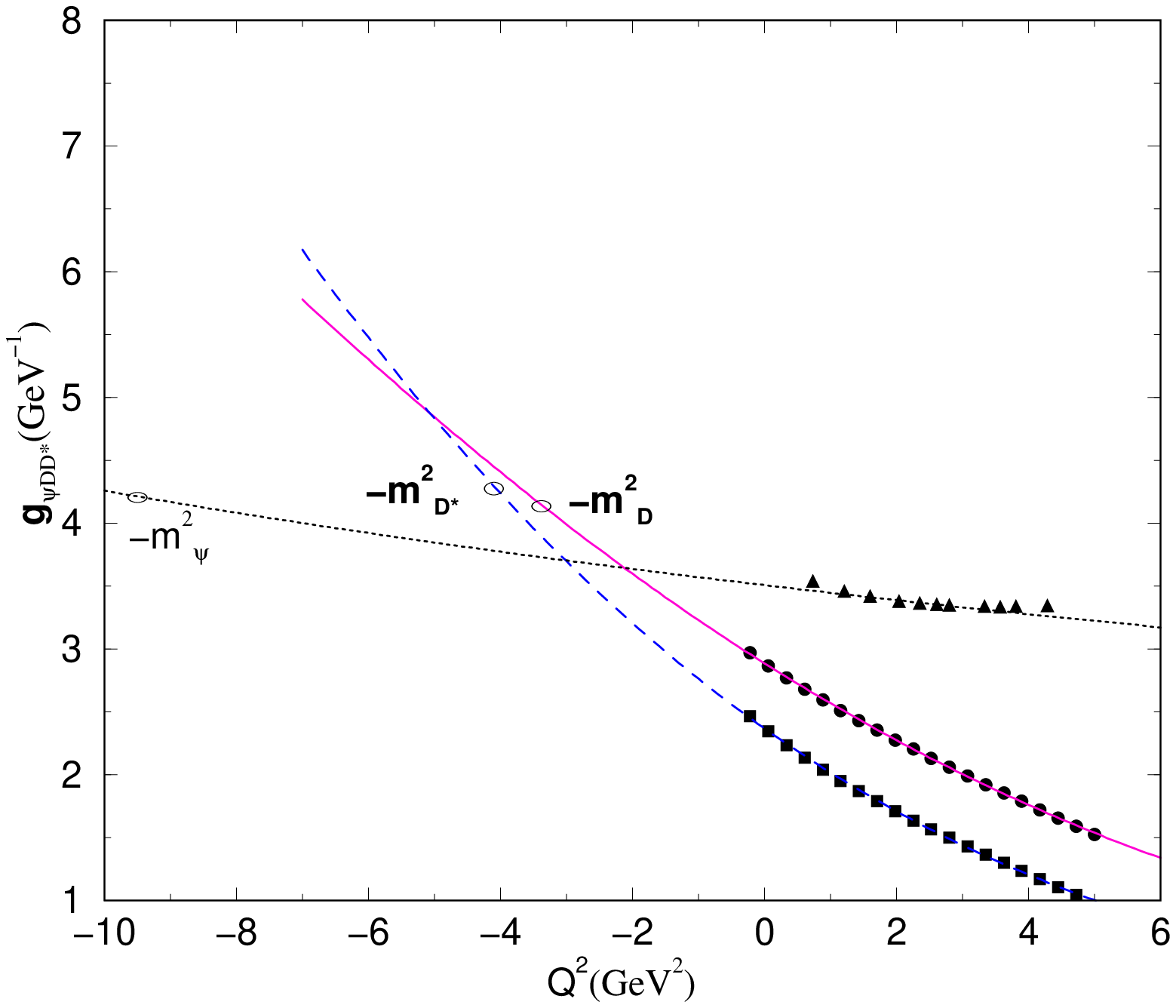}}
\end{center}
\protect\caption{Momentum dependence of the $J/\psi DD^*$ form factors. The 
dotted, dashed and solid lines give the parametrization of the QCDSR results 
(triangles, squares and circles).}
\label{f_psidds4.eps}
\end{figure}

As another interesting example, we consider the vertex $J/\psi D D^*$. Here we try to 
improve the procedure described above, calculating three form factors (one for each 
off-shell particle). This new procedure imposes a more stringent condition on the 
parametrizations.  Fixing $M^2$ and $\mli$ to the values of the incoming and outgoing
meson masses we  show, in Fig.~\ref{f_psidds4.eps}, the momentum dependence of the QCDSR 
results for the three form factors 
$g_{J/\psi D D^*}^{(D)}$, $g_{J/\psi D D^*}^{(D^*)}$ and $g_{J/\psi D D^*}^{
(J/\psi)}$ through the circles, squares and triangles respectively.
Since our  approach 
cannot be used at $Q^2 \ll 0$, in order to extract the $g_{J/\psi D D^*}$
coupling from the form factors we  extrapolate the curves to 
the mass of the off-shell meson, shown as open circles in Fig.~\ref{f_psidds4.eps}. 
In order to do this extrapolation we fit 
the QCDSR results  with an analytical expression. We tried to fit
our results to a monopole form, since this is very often used 
for form factors, but the fit was only good for $g_{J/\psi D D^*}^{(J/\psi)}$. 
For $g_{J/\psi D D^*}^{(D)}$ and $g_{J/\psi D D^*}^{(D^*)}$
we obtained  good fits using  a Gaussian form. 
These fits are also shown in Fig.~\ref{f_psidds4.eps} 
through the dotted, solid and dashed  
lines respectively. From Fig.~\ref{f_psidds4.eps} 
we see that all three form factors lead to compatible values for the coupling 
constant when extrapolated to the off-shell meson masses (open circles in 
Fig.~\ref{f_psidds4.eps}).

From the parametrizations  we can also extract the cutoff parameter,
$\Lambda$, associated with the form factors. The general expression for
the Gaussian parametrization is: $A\exp{[-(Q^2+B)^2/\Lambda^4]}$, which 
gives $\Lambda\sim 4.5~\GeV$ when the off-shell meson is $D$ or $D^*$. For
the monopole parametrization the general expression is: $g[(\Lambda^2-m^2)/(
\Lambda^2+Q^2)]$. Therefore, for an off-shell $J/\psi$ we get
$\Lambda\sim7.5~\GeV$. It is very interesting to notice that the value of the 
cutoff is  directly associated with the mass of the off-shell meson in
the vertex. The form factor is harder (i.e., the curve in the Figure is  flatter) 
if the off-shell meson is heavier.

\subsection{Hadronic loops}

Coming back to the $D^* D \pi$ vertex, we observe that, 
while the obtained number,  $g_{D^* D \pi} = 14.0 \pm 1.5$,  is not far from the 
experimental value, 
there is still a discrepancy. The procedure of fitting the QCDSR points in the deep 
euclidean region and extrapolating them to the time-like region  contains 
systematic uncertainties associated to   {\it   the analytical form chosen for the 
parametrizations}, i.e., monopole, exponential or gaussian. We tried to reduce this 
systematic uncertainty performing a double (and also a triple as in the case of the 
$J/\psi D^* D $ vertex) fit.  However it would be desirable to have a more physical way 
to reduce the systematic uncertainty.

In \cite{mane} the authors revisited  this problem, employing  hadronic loops, calculated by 
means of effective field theories (EFT), in order to produce a better parametrization 
for $D^* D \pi$ results calculated with  QCDSR.
Purely hadronic calculations are independent from QCDSR and involve the choice of an 
effective Lagrangian, 
including the possible requirements of chiral symmetry and/or SU(4). Beyond the tree 
level, one has to deal with 
the problems and uncertainties associated with renormalization. As it was discussed in 
\cite{mane} a suitable combination of  EFT and QCDSR results allows the reduction  of 
undesired indeterminacies of both approaches, improving  their predictive power.

\begin{figure}[h]
\begin{center}
\epsfxsize=9cm
\leavevmode
\hbox{\epsffile{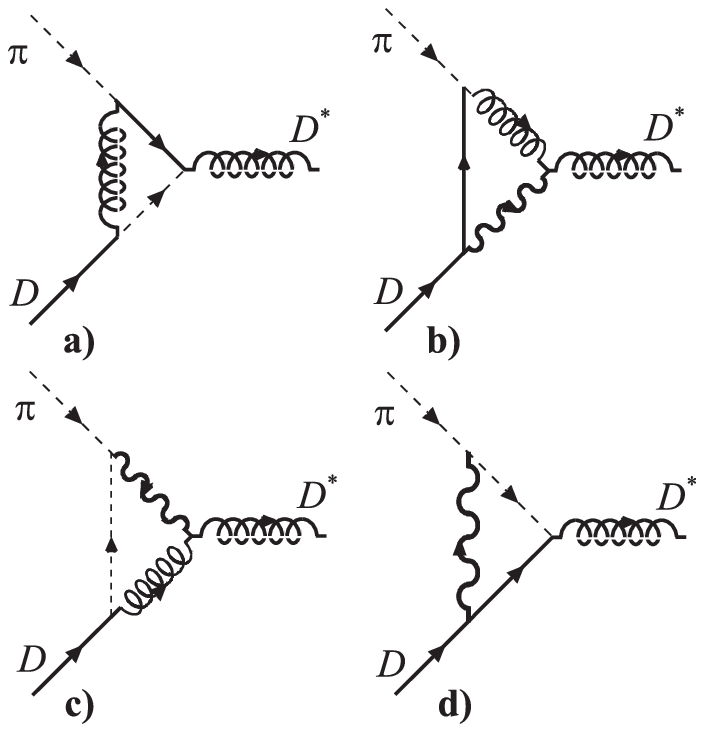}}
\end{center}
\protect\caption{Meson loop contributions to the $D^* D \pi$ form factor:
``triangle'' diagrams. In the internal triangles  the solid, dashed, wavy and spring lines
represent  a  virtual $D$, $\pi$,  $\rho$ and  $D^*$ respectively.}
\label{manefig2}
\end{figure}

The full $D^* D \pi$ vertex function in a hadronic approach involves the computation of 
several diagrams. Leading contributions to this vertex come from both the tree interaction 
and from the  
diagrams depicted in Fig. \ref{manefig2}.
Meson loops are a necessary consequence of quantum field theory and do contribute to several 
hadronic observables. 
In practice, due to problems associated with infinities, renormalization becomes unavoidable 
in the evaluation of loop corrections to observables. 
Here the basic idea is to isolate the unknown loop parameters into some basic constants, in 
such a way that they can be determined by matching the results of  loop and QCDSR results. 

We  note that some diagrams, such as, for instance,  that in Fig. \ref{manefig2} a, 
contain internal vertices 
involving the $D^* D \pi$ coupling.  This suggests that the calculation is ``cyclic'', since one needs to use the 
$D^* D \pi$  form factor in order to calculate the $D^* D \pi$  form factor. 
Actually, there are differences between the internal particles and the external ones. 
The former are always virtual, whereas the latter may be either real or put on mass shell in the extraction of the 
coupling constant.  In the framework of perturbation theory, at leading order, internal particles are treated as 
elementary, without structure.
They are assumed to be point-like and the evaluation of leading terms does not require the use of internal form factors. 
Consistently, one must use bare coupling constants for these interactions.

Since there are heavy mesons circulating in the loops shown in Fig. \ref{manefig2}, 
one might  argue that other states should also be included. 
We do have, for example, fermion-antifermion components such as $\bar{N}N$ or $\bar{\Lambda}_c\Lambda_c$ in the loops.
An incoming positive pion can split into a $p$ plus a $\bar{n}$, and so on.
However, in a different context \cite{speth},  
it has been shown that this kind of splitting is suppressed with respect 
to the pion $\rightarrow$ meson-meson splitting, by one order of magnitude. 
The neglect of this kind of contribution seems therefore justified.
The same holds for the possibility of strangeness circulating in the loop, associated with  virtual states 
such as $D_s$, 
$D^*_s$, $K$ and $K^*$. 
Using only $\pi$'s, $\rho$'s, $D$'s and $D^*$'s  the  low and  high $Q^2$ 
regions of the form factor are covered. 
Thus it is enough  to work with a simple effective theory.
It is convenient to use  the effective Lagrangian (\ref{dsdpi}), which is constrained by  
SU(2) flavor and chiral symmetries, as well as gauge invariance.  
The  coupling constant appearing in the Lagrangian is the  bare one.
The $\r$ couplings are assumed to be universal and are implemented by covariant 
derivatives of the form $ \cD^\m = \d^\m - i g_\r \;\bT\cd \bro^\m\ $, where 
$g_\r$ is the universal coupling constant and 
$\bT$ is the isospin matrix suited to the field  upon which the derivative $\cD^\m$  acts.   
With the proper Lagrangians it is possible to  write and evaluate all the contributions  to 
the total vertex function.
\begin{figure}[h]
\begin{center}
\epsfxsize=9cm
\leavevmode
\hbox{\epsffile{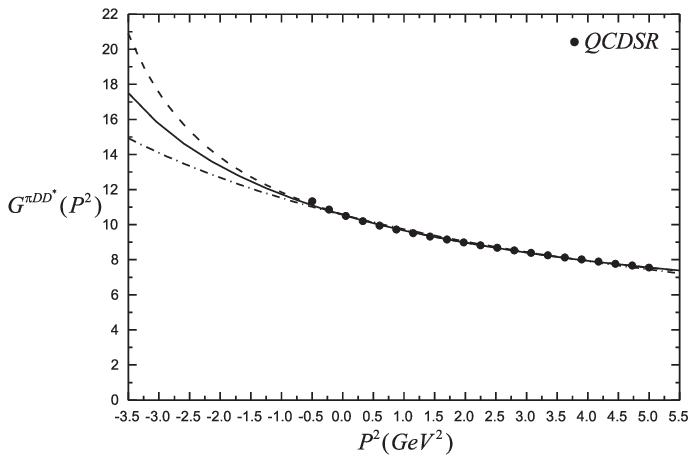}}
\end{center}
\protect\caption{The $D^* D \pi$ form factor. Dots: QCDSR from \protect\cite{02r}; 
solid, dash and dash-dotted lines are fits obtained with eq. (\protect\ref{gfinal}),  
(\protect\ref{GFI}) and (\protect\ref{GFII}), respectively.}
\label{manefig5}
\end{figure}

The $\pi (q)\;D(p)\; D_\a^*(p')$ vertex function $\Gamma_{\mu} (p^2)$ for an off-shell 
$D$ is written as:
\bea
\G_{\mu} (p^2) = - \; q_{\mu} \,  g^{(D)}_{D^* D \pi} (p^2) ,  
\label{3.1}
\eea
where $g^{(D)}_{D^* D \pi} (p^2)$ is the  form factor, such that the physical coupling constant 
is $g_{D^* D \pi} = g^{(D)}_{D^* D \pi}( m_D^2)$.
In \cite{mane}  two kinds of  loop corrections  to this vertex were considered, 
containing pion and $\rho$ intermediate states, denoted respectively by $F_\p(p^2)$ 
and $F_\r(p^2)$.
The perturbative evaluation of these functions gives rise to divergent integrals and 
$g^{(D)}_{D^* D \pi}(p^2)$ can be determined only up to yet unknown renormalization constants. 
The use of standard loop integration techniques, such as dimensional regularization and 
$\overline{MS}$ subtraction of divergences, for all diagrams, allows one to write the form 
factor as:
\bea
g^{(D)}_{D^* D \pi}(p^2) = K + C_\p \;F_\p( p^2) + C_\r \;F_\r (p^2) \;,
\label{gfinal}
\eea
where $K$, $C_\p$ and $C_\r$ are constants.
These constants incorporate the bare couplings, the usual parameters associated with 
renormalization  and here they are determined by comparing  $g^{(D)}_{D^* D \pi}(p^2)$ with the results 
from QCD sum  rules.  Keeping only the terms which depend on $p^2$   the explicit evaluation of the 
diagrams can be performed and  the form factor as a function of $p^2$, the $D$ four-momentum squared, can be 
obtained. At this stage, it still contains three unknown parameters, which are determined by 
adjusting the function $g^{(D)}_{D^* D \pi}(p^2)$ to the QCD sum rule points taken from  \cite{02r}.
Those results are displayed in Fig. \ref{manefig5}, where $P^2 = -p^2$, together with 
the best  fit ($\chi^2\sim 10^{-3}$) represented by the solid line.
Computing the value of $g^{(D)}_{D^* D \pi}(p^2)$ at $p^2=m_D^2$, one arrives at the following value for 
the coupling constant:
\beq
g_ {D^* D \pi} = 17.5 \pm 1.5 \;,
\label{Dresult}
\eeq
in very good  agreement with experiment. The errors quoted come from the QCDSR points, 
which contain a typical error of $\simeq$ 10 \%.
In the same Figure we can also see  the results of the fits of the QCDSR points with 
two mixed monople-dipole structures with three free parameters, namely:
\bea
g^{I}(p^2) &=& C \lb \frac{\L_1^2-m_D^2}{\L_1^2-p^2} + \lp \frac{\L_2^2-m_D^2}
{\L_2^2- p^2}\rp^2 \rb \;,
\label{GFI}\\[2mm]
g^{II} (p^2) &=& C_1 \, \frac{\L^2-m_D^2}{\L^2-p^2} + C_2\,\lp \frac{\L^2-m_D^2}
{\L^2- p^2}\rp^2  \;,
\label{GFII}
\eea
which yield $\chi^2_I\sim 10^{-3}$ (dashed line) and $\chi^2_{II}\sim 10^{-2}$ (dash-dotted 
line), respectively. Looking at  Fig. \ref{manefig5} we learn that these alternative 
structures, reasonable as they  are, diverge significantly
from the loop calculation  in the region where the $D$ is not too off-shell, stressing
the importance of a proper hadronic treatment of the form factor in that region.

As far as practical applications are concerned, our numerical results for the  form factor 
$g_{D^* D \pi}(p^2)$, in the whole range $- m^2_{D} \leq P^2 < 5$ GeV$^2$,
are very well described by the mixed monopole-dipole structure given by Eq. (\ref{GFI})
with the parameters $C = 8.7$, $\Lambda_1=5.1$ GeV and  $\Lambda_2=2.9$ GeV.
These results suggest that the use of meson loops can  reduce the uncertainty in the 
extrapolation of form factors, computed in the space-like region by means of QCDSR, to the 
time-like region, with the corresponding increase in the reliability of predictions for 
coupling constants.  Apart from the approximations described above,  the  procedure has no 
new source of errors.

To conclude this section, we have discussed a new method of improving QCDSR calculations 
of hadronic form factors, which 
consists in matching  QCDSR results, valid mainly in the deep euclidean region, to meson 
loop calculations,
valid when the $D$ is not too off-shell.
This matching is well justified from the physical point of view,  since in the intermediate 
and  large $Q^2$ regions
the relevant degrees of freedom are the quarks and gluons,  with non-perturbative 
corrections taken into 
account through the QCD condensates.  
The opposite happens  for low values of $Q^2$, where sum rules calculations become 
non-reliable due to the lack 
of a large mass scale. 
At this point, the meson exchange dynamics  becomes a reliable tool,
but it depends on unknown constants associated with the renormalization of the mesonic vertices.  
Although the exact frontier between  meson dynamics and  QCDSR cannot be precisely known,
the success of the method in the example  considered here supports the view that the matching
may become useful in increasing the predictive power of both procedures.

\section{Results}
\label{sec_res}

\subsection{Form factors and couplings}

We have applied the procedure described in the previous sections to the following vertices 
$D^* D \pi$, $D^* D^* \pi$, $D D \rho$, 
$D^* D \rho$, $D^* D^* \rho$, $J/\psi D D$, $J/\psi D^* D$ and $J/\psi D^* D^*$. 
As mentioned above, for each vertex 
we obtain two sets of points, which have been parametrized by the following forms: 
\beqa
(I) \,\,\,\,\,\,\,\,  g^{(M_i)}_{M_1 M_2 M_3} &=& \frac{A}{Q^2 + B}   \label{mono} \\
(II) \,\,\,\,\,\,\,\,  g^{(M_i)}_{M_1 M_2 M_3} &=& A \exp[-(Q^2/B)] \label{exp}  \\
(III) \,\,\,\,\,\,\,\,  g^{(M_i)}_{M_1 M_2 M_3} &=& A \exp[-(Q^2+C)^2/ B] . \label{gauss}
\eeqa
In Table \ref{tab1} each line refers to the vertex indicated in the first column. 
In the second, third and fourth columns we present the values of the parameters  
A and B for the case where a heavier meson in the vertex ($M_1$) is off-shell, 
indicating also which parametrization ((I), (II) or (III)) was employed.  In the 
fifth, sixth, seventh and eighth columns we show the parameters and type of 
parametrization used in the case where a lighter meson ($M_2$) is off-shell. 
\begin{table}[h]
\centering
\begin{tabular}{lrrrrrrrr}
\hline
$M_1 \, M_2 \, M_3$  & 
  & 
  & $M_1$ off
&$\,\,\,\,\,\,\,\,\,\,\,\,\,\,\,\,\,\,\,\,\,\,\,\,\,$  &  &  
& $M_2$   off
&  \\
 & Form & A & B & $\,\,\,\,\,\,\,\,\,\,\,\,\,\,\,\,\,\,\,\,\,\,\,\,\,$  & Form  & A & B & C\\
\hline
$D \pi D^*$ & (I) & 126 & 11.9 & $\,\,\,\,\,\,\,\,\,\,\,\,\,\,\,\,\,\,\,\,\,\,\,\,\,$ & (II) & 15.5 & 1.48 &\\
$J/\psi  D D^*$  & (I) & 200 & 57 & $\,\,\,\,\,\,\,\,\,\,\,\,\,\,\,\,\,\,\,\,\,\,\,\,\,$  & (III) & 13 & 450 & 26\\
$J/\psi D D$  & (I) & 306 & 63 &$\,\,\,\,\,\,\,\,\,\,\,\,\,\,\,\,\,\,\,\,\,\,\,\,\,$ & (III)& 15 & 250 &  20\\
$D \rho D$  & (I) & 37.5 & 12.1 & $\,\,\,\,\,\,\,\,\,\,\,\,\,\,\,\,\,\,\,\,\,\,\,\,\,$  & (II)& 2.5 & 0.98& \\
$J/\psi D^* D^*$ & (I) & 400 & 78.5& $\,\,\,\,\,\,\,\,\,\,\,\,\,\,\,\,\,\,\,\,\,\,\,\,\,$   & (II) & 1.96 & 3.5 &\\
$D^* \rho D^*$ & (II) & 4.9 & 13.3 & $\,\,\,\,\,\,\,\,\,\,\,\,\,\,\,\,\,\,\,\,\,\,\,\,\,$ & (II)& 5.2 & 2.7 & \\
$D^* \pi D^*$ & (II) & 4.8 & 6.8 & $\,\,\,\,\,\,\,\,\,\,\,\,\,\,\,\,\,\,\,\,\,\,\,\,\,$ & (II)& 8.5 & 3.4 & \\
$D \rho D^*$   & (I) & 234 & 44  & $\,\,\,\,\,\,\,\,\,\,\,\,\,\,\,\,\,\,\,\,\,\,\,\,\,$ & (II)& 5.1 & 4.3 & \\
\hline
\end{tabular}
\caption{Parameters used in (\ref{mono}), (\ref{exp}) and (\ref{gauss}). B and C are 
in GeV$^2$ and A is  either in GeV or  GeV$^2$, depending on the vertex. All the isovector
mesons $\pi$ and $\rho$ appearing in the Table are charged.}
\label{tab1}
\end{table}
The numbers presented in Table \ref{tab2} summarize our results. They contain uncertainties  coming 
from different sources, which will be discussed in the next section.  The vertex coupling 
constants can be obtained from the form factors and they are presented in Table \ref{tab2}.  
A  comparison with other estimates is going to be done in section \ref{sec_disc}.

\begin{table}
\centering
\begin{tabular}{ll}
\hline
$M_1 \, M_2 \, M_3$ &
$g_{M_1 \, M_2 \, M_3}$\\
\hline
$D \pi D^*$ & $9.9 \pm 1.0$  \\
$J/\psi  D D^*$ & $4.0 \pm 0.6$  GeV$^{-1}$ \\
$J/\psi D D$ & $5.8 \pm 0.9$ \\
$D \rho D$ &  $3.0 \pm 0.2$ \\
$J/\psi D^* D^*$ &  $6.2 \pm 0.9$ \\
$D^* \rho D^*$ & $4.7 \pm 0.2$ \\
$D^* \pi D^*$ &  $6.1 \pm 0.7$  GeV$^{-1}$ \\
$D \rho D^*$ &  $4.3 \pm 0.9$  GeV$^{-1}$ \\
\hline
\end{tabular}
\caption{Coupling constants for neutral isovector mesons ($\pi^0$ and $\rho^0$). }
\label{tab2}
\end{table}

\subsection{Uncertainties}

\begin{figure}[h]
\begin{center}
\epsfxsize=11cm
\leavevmode
\hbox{\epsffile{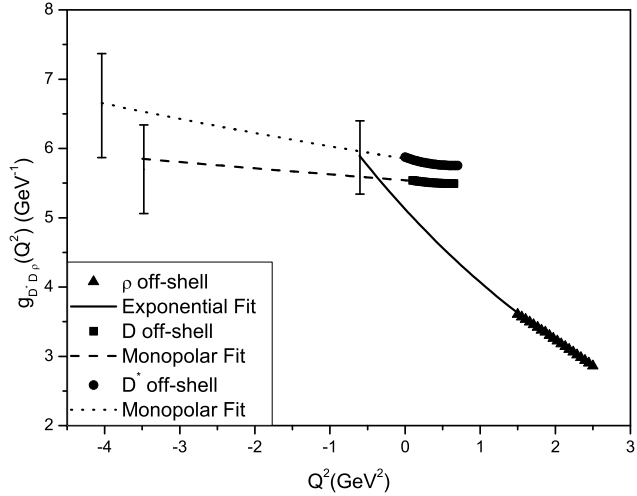}}
\end{center}
\protect\caption{The $D^* D \rho$ form factor. 
$g^{(D)}_{D^{*}D \rho}$ (squares),  $g^{(\rho)}_{D^{*}D \rho}$ (triangles) and
$g^{(D^*)}_{D^{*}D \rho}$ (circles)  form factors as a function of $Q^2$.
The dotted, solid an dashed lines correspond to the  parametrizations discussed in the
text. The vertical bars show the theoretical errors in the coupling constants  once all 
variations in the parameters are taken into account, as explained in the text.}
\label{brunofig5}
\end{figure}

We  consider now, as an example,  the vertex $D^* D \rho$ and discuss, one by one, all the sources 
of uncertainties in the  calculation of the form factors.  The sum rules for $D$, $\rho$  
and  $D^{*}$ off-shell are given by:
$$
C  \frac{ g_{D^{*} D \rho}^{(D)}}{(Q^2 + m_{D}^2)} e^{-\frac{m_{\rho}^2}{\mli}} e^{-\frac{m_{D^{*}}^2}{M^2}} 
= - \frac{1}{4\pi^2} \int ds  \int du~ \rho^{(D)}(u,s,t) \, e^{-\frac{s}{M^{2}}} e^{-\frac{u}{\mli}} \nonumber  
$$
\beq
- \, \langle \bar q q \rangle \,  e^{- m_c^2/M^2} , 
\label{s1}
\eeq
\beq
C \frac{ g_{D^{*} D \rho}^{(\rho)}}{(Q^2 + m_{\rho}^2)} e^{-\frac{m_{D}^2}{\mli}} e^{-\frac{m_{D^{*}}^2}{M^2}}
= - \frac{1}{4\pi^2} \int ds  \int du~ \rho^{(\rho)}(u,s,t) \, e^{-\frac{s}{M^{2}}}e^{-\frac{u}{\mli}} ,
\label{s2}
\eeq
and 
$$
C \frac{  g_{D^{*} D \rho}^{(D^*)}}{(Q^2 + m_{D^*}^2)} e^{-\frac{m_{D}^2}{\mli}} e^{-\frac{m_{\rho}^2}{M^2}}
= - \frac{1}{4\pi^2} \int ds  \int du~ \rho^{(D^*)}(u,s,t) \, e^{-\frac{s}{\mli}} e^{-\frac{u}{\mli}}  \nonumber
$$
\beq
 + \,  \langle \bar q q \rangle \, e^{- m_c^2/M'^2} ,
\label{s3}
\eeq
where $t = - Q^2$ and  the functions $\rho^{(D)}$, $\rho^{(\rho)}$ and $\rho^{(D^*)}$ are the double discontinuities associated to the 
perturbative diagram of the sum rules with an off-shell $D$, $\rho$ and $D^*$ off-shell respectively. They are given by:
\beq
\rho^{(D)}(u,s,t)= \frac{3m_c}{\sqrt{\lambda}} \left[\frac{u (2 m_c^2 -s -t +u)} {\lambda}\right] ,
\label{rhodoff}
\eeq
with $\lambda =(u+s-t)^2 -4us$. The integration limits in (\ref{s1}) are:
$$
0 \, < \, u \, < \, \frac{m_c^2 (s + t - m_c^2) - st}{m_c^2} ,
$$
and
$$
m_c^2 \, < \, s \, < \, s_0 ,
$$
The perturbative contribution for the double discontinuity for an off-shell $\rho$ meson is given by:
\beq
\rho^{(\rho)}(u,s,t)=
\frac{3m_c \, t}{\lambda^{3/2}} \left[u + s - t - 2 m_c^2 \right] ,
\label{rhorhooff}
\eeq
and the corresponding integration limits in (\ref{s2})  are:
$$
\frac{m_c^2 (s  - t  - m_c^2) }{s - m_c^2}  \, < \, u \, < \, u_0 ,
$$
and
$$
m_c^2 \, < \, s \, < \, s_0 ,
$$
The perturbative contribution for the double discontinuity for an off-shell $D^*$ meson is given by:
\beq
\rho^{(D^*)}(u,s,t)= \frac{3m_c}{{\lambda}^{3/2}} \left[s (2 m_c^2 + s -t - u)\right] ,
\label{rhodsoff}
\eeq
The integration limits in  (\ref{s3}) are:
$$
t \, < \, u \, < \, \frac{m_c^2 ( t -s - m_c^2)}{t-m_c^2} ,
$$
and
$$
0 \, < \, s \, < \, s_0 ,
$$
In the above expressions $C$ is a constant defined as:
$$
C={m_D^2f_D\over m_c}m_{\rho}f_{\rho}m_{D^{*}}f_{D^{*}} ,
$$
where $f_D$, $f_{\rho}$ and  $f_{D^*}$ are the  decay constants of the mesons $D$, $\rho$ and 
$D^*$ respectively. 
The numerical evaluation of the above expressions is done with the numerical inputs shown in
Table \ref{table1} and the resulting points are shown in Fig. \ref{brunofig5}.
The triangles, squares and circles are the results  
for the $g_{D^{*}D \rho}^{(\rho)}(Q^2)$, 
$g_{D^{*}D \rho}^{(D)}(Q^2)$ and $g_{D^{*}D \rho}^{(D^*)}(Q^2)$ form factors respectively.  
As indicated in Table \ref{tab1},  in the case of an off-shell $ D $ meson, our numerical 
results can be fitted by the following monopolar parametrization (shown by the dashed  
line  in Fig.~\ref{brunofig5}):
\begin{equation}
g_{D^{*}D \rho}^{(D)}(Q^2)= \frac{234} {Q^2 + 44} ,
\label{monodoff}
\end{equation}
where the function $g_{D^{*}D \rho}^{(D)}(Q^2)$ has the units of GeV$^{-1}$. Following the 
procedure discussed above, we define the coupling constant as the value of the 
form factor at $Q^2= -m^2_{M}$, where $m_{M}$ is the mass of the  meson $M$. 
Therefore, using $Q^2=-m_{D}^2$ in Eq~(\ref{monodoff}), the resulting coupling 
constant is $g^{(D)}_{D^{*}D \rho } =   5.76 \, \GeV^{-1}$.
For an off-shell $\rho$ meson  our sum rule results  can  be 
fitted by an exponential parametrization, which is represented by the 
solid  line in Fig.~\ref{brunofig5}:
\begin{equation}
g_{D^{*}D \rho }^{(\rho)}(Q^2)= 5.1 \,  e^{-Q^2/4.3}  .
\label{exprhooff}
\end{equation}
Using $Q^2=-m^2_{\rho}$ in Eq~(\ref{exprhooff}) we get 
$g^{(\rho)}_{D^{*}D \rho }= 5.89  \,  \GeV^{-1}$.
In the case of an off-shell $ D^* $ meson, our numerical results can be
fitted by the following monopolar parametrization (shown by the dotted line in
Fig.~\ref{brunofig5}):
\begin{equation}
g_{D^{*}D \rho}^{(D^*)}(Q^2)= \frac{195.8} {Q^2 + 33.5} .
\label{monodsoff}
\end{equation}
Evaluating  this form factor at  $Q^2=-m^2_{D^*}$ we find the coupling  
$g^{(D^*)}_{D^{*}D \rho }= 6.65  \,  \GeV^{-1}$.

Looking at Fig. \ref{brunofig5}  we can observe that the $D$ off-shell form fator is 
much harder  than the $\rho$ off-shell 
one.  This agrees 
with the behavior observed in Figs. \ref{psidsds_struc},   \ref{fig2} and \ref{f_psidds4.eps}:   
the heavier is the off-shell meson, the harder is its form factor.  Following this same trend, 
we would expect  the  $D^*$ off-shell form factor to be even  harder than the  $D$ off-shell one.  
However, comparing the dashed and dotted lines in   Fig. \ref{brunofig5},   this seems not to be 
the case: the slope of the $D^*$ curve is slightly bigger than the one of the $D$ curve.  Since 
their mass difference  is relatively small ($ \simeq 150 $ MeV) the two curves should have  almost the 
same slope.  The observed difference is an indication of the limited precision of our method.

The form factors (\ref{monodoff}), (\ref{exprhooff}), (\ref{monodsoff}) and their 
extrapolations to the on-shell points leading to the coupling constants do not contain 
error bars. In fact, a careful and systematic study of errors in QCDSR calculations is hard 
to find in the literature. We took Refs. \cite{derek} as a guide. 
In Fig. \ref{brunofig5} we can see the theoretical error bars at the endpoints of the three 
curves. In what follows we describe how we obtain them.  
First, we compute the sum rules (\ref{s1}), (\ref{s2}) and (\ref{s3}) 
extensively, taking into account the errors in the masses, decay constants, condensates,
choice of the Borel mass and continuum threshold parameters. 
In each computation all the parameters are kept fixed, except one, which is  changed 
according to its intrinsic error.  The errors in the quark and gluon condensates, in the masses and 
decay constants  are listed in Table \ref{table1}.  The three Borel masses were chosen 
in the interval $2.7 \leq M^2 \leq 3.3$ GeV$^2$ for an off-shell $\rho$  and an off-shell 
$D^*$ and in the interval $27 \leq  M^2 \leq 33 $ GeV$^2$ for an off-shell $D$.
After each round of calculation of the three sum rules, we obtain three sets of points 
which are then fitted and extrapolated to the respective on-shell points.  The sets of 
points  are all fitted with the forms  (\ref{monodoff}), (\ref{exprhooff}) and  
(\ref{monodsoff}),  but
for each set the numerical constants appearing in these forms are different. 

Every extrapolation introduces some 
ambiguity in the final results, since we have the freedom to fit a set of points with 
different parametrizations. In our case  this freedom is strongly reduced because we require 
that all the three parametrizations lead to approximately the same coupling constant. 
In Fig. \ref{brunofig5}  this requirement forces the three endpoints of 
(\ref{monodoff}), (\ref{exprhooff}) and  (\ref{monodsoff}), which are taken at the squared 
masses of the corresponding  particles, to coincide, i.e., to have approximately the same 
height in the 
figure. Of course, due to the approximations used, we can not expect this matching to be 
perfect.  Once this procedure is completed  and we determine the three coupling constants
with an error corresponding solely to the variation of one parameter, we move to the next 
parameter to be varied, keeping all others fixed and repeat the procedure.  In each step we 
can have an idea of how sensitive is each coupling constant to the parameter under 
consideration.  In the end, for each coupling constant we take the average of all encountered 
values and calculate also the global error, which is shown in  Fig.  \ref{brunofig5}  as an error 
bar at the on-shell point. The final number is then obtained taking the average of the 
three couplings found and the final error is also obtained from the  errors  of each coupling. 

Among the sources of errors, one deserves a special discussion. Very often in QCDSR 
calculations, appreciable uncertainties in the results come from the lack of knowledge on 
the continuum threshold parameters.  In order to study the dependence of our results on 
these parameters, we vary 
$\Delta_{s,u}$ between 
$0.4 \GeV \le \Delta_{s,u} \le 0.6 \GeV$ in the  sum rule (\ref{s2}),  between 
$0.4 \GeV \le \Delta_{s} \le 0.6 \GeV$ and $0.65 \GeV \le \Delta_{u} \le 0.75 \GeV$
in the sum rule (\ref{s1}) and  between 
$0.65 \GeV \le \Delta_{s} \le 0.75 \GeV$ and $0.50 \GeV \le \Delta_{u} \le 0.70 \GeV$
in the sum rule (\ref{s3}). 
This variation produces new sets of curves which are shown in 
Fig.~\ref{dsdrho_ex} and  give us an uncertainty range in the resulting coupling constants 
$g_{D^{*}D \rho }^{(D)}$ and $g_{D^{*}D \rho }^{(\rho)}$. For the sake of clarity we did 
not include the lines corresponding to the coupling  $g_{D^{*}D \rho }^{(D^*)}$. 
Surprisingly, in the case of the form factor  $g_{D^{*}D \rho }^{(\rho)}$, we observe 
a convergence of the extrapolation lines, which reduces the final error. Due to 
this accident, the continuum threshold parameters are not, in the  
$g_{D^{*}D \rho }^{(\rho)}$  case, the ultimate source of error. Their contribution 
(denoted by  $\Delta$ in the Tables) is still  significant, as it can be seen in 
Tables \ref{table2}, \ref{table3} and \ref{table4}, but now they have less impact on the final 
error than the uncertainties in the decay constants $f_D$ and $f_{D^*}$ and in the charm quark 
mass. In these Tables we show in  the first column 
the quantity which was varied, in the second  the average coupling constant resulting from 
that variation, in the third the standard deviation and in the fourth the percentual 
significance of $\sigma$. 

After scanning the space of reasonable values of all the parameters, we conclude that, in
spite of the inherent uncertainties, the sum rules really point to a value of the coupling 
constant!  Of course, as in most of QCDSR calculations, the lack of precision is due to 
the ``usual suspects'', i.e., continuum threshold parameters, decay constants, heavy quark 
masses and condensates. A comparison of the Tables shows an intriguing aspect, namely  
that some of  the input quantities affect  each of the three sum rules in a quite different 
way. This may be a signal that some of the sum rules are less reliable than others. A 
deeper investigation of this question would involve several refinements, such as the 
calculation of $\alpha_s$ corrections and higher order terms in the OPE.

\begin{table}
\centering
\begin{tabular}{cccc}
\hline
Quantity & $ \left\langle g^{(\rho)}_{D^* D_{\rho}} \right\rangle $ & $  \sigma $ & $\sigma \, (\%) $ \\
\hline
$ \Delta $  & 5.86 & 0.08 & 1.4 \\
$ f_\rho $ & 5.89 & 0.01 & 0.1\\
$ f_D $ & 5.41 & 0.65 & 12.0\\
$ f_D* $ & 5.90 & 0.40 & 6.8\\
$ M^2 $ & 5.90 & 0.10 & 1.7\\
$ m_c $  & 5.97 & 0.40 & 7.4\\
\hline
\end{tabular}
\caption{Changes in $g^{(\rho)}_{D^* D \rho }$ induced by changes in different quantitities.}
\label{table2}
\end{table}

\begin{table}
\centering
\begin{tabular}{cccc}
\hline
Quantity & $ \left\langle g^{(D)}_{D^* D \rho } \right\rangle $ & $ \sigma $ & $\sigma \,  (\%) $ \\
\hline
$ \Delta $  & 5.95 & 0.87 & 14.7 \\
$ f_\rho $ & 5.76 & 0.01 & 0.1\\
$ f_D $ & 5.30 & 0.64 & 12.0\\
$ f_D* $ & 5.80 & 0.40 & 6.8\\
$ M^2 $ & 5.76 & 0.05 & 0.8\\
$ m_c $  & 5.70 & 0.30 & 5.6\\
$\langle \bar q q\rangle $ & 5.77 & 0.04 & 0.8\\
\hline
\end{tabular}
\caption{Changes in $g^{(D)}_{D^* D \rho}$ induced by changes in different quantitities.}
\label{table3}
\end{table}
\begin{table}
\centering
\begin{tabular}{cccc}
\hline
Quantity & $ \left\langle g^{(D^*)}_{D*D \rho} \right\rangle $ & $ 
\sigma $ & $\sigma \,  (\%) $ \\
\hline
$\Delta$    & 7.00   & 1.00    & 14.3   \\
$f_\rho$    & 6.61   & 0.07    & 1.1    \\		
$f_D$	    & 6.11   & 0.74    & 12.0   \\
$f_{D^*}$   & 6.69   & 0.46    & 6.8    \\
$M^2$	    & 6.65   & 0.19    & 2.8   \\
$m_c$	    & 6.61   & 0.06    & 0.8    \\
$\langle \bar q q\rangle $ & 6.66 & 0.08 & 1.3\\
\hline
\end{tabular}
\caption{Changes in $g^{(D^*)}_{D*D \rho}$ induced by changes in different quantitities.}
\label{table4}
\end{table}

Considering the results presented in the Tables, the couplings  are:
$$
g_{D^{*}D \rho }^{(D)}  = 5.71 \pm 0.62  \, \GeV^{-1} ,
$$
$$
g_{D^{*}D \rho }^{(\rho)} = 5.87 \pm 0.53  \, \GeV^{-1} ,
$$
and
$$
g^{(D^*)}_{D^{*}D \rho }= 6.63     \pm 0.73  \,  \GeV^{-1} ,
$$
We can see that the three cases considered here, off-shell $D $, $\rho $ and $D^*$,
give compatible results for the coupling constant. Considering all the uncertainties
and taking the average between the obtained values we have:
\begin{equation}
g_{D^{*}D \rho }=  \Big (6.1 \pm 1.3  \Big ) \,  \GeV^{-1} ,
\label{finalcoupling}
\end{equation}
Our results were obtained for certain  concrete choices of currents,  which represent 
charged states. Consequently,  the
obtained couplings are for charged states. As it will be discussed in the next section, 
from the coupling of charged states 
we can get the  coupling  of neutral $\rho$  states, which will be identified with the 
``generic coupling'',  through the relation:
\begin{equation}
g_{D^* D \rho }  = g_{D^* D \rho^0 } =  \frac{g_{\rho^+ D^{0} D^{*+}}}{\sqrt{2}}
= \frac{g_{\rho^- D^{0} D^{*+}}}{\sqrt{2}} .
\end{equation}
Therefore the final value of the coupling constant listed in Table \ref{tab2}  is:
$$ 
g_{D^{*}D \rho }= \Big ( 6.1  \pm 1.3 \Big ) /\sqrt{2}=
\Big (4.3 \, \pm 0.9 \Big ) \, \GeV^{-1}  ,
$$
To close this section we emphasize that, from the analysis of the Tables we 
conclude that, in the present context, the average error of our calculations is in the 
range from 10 to 15 \%.

\section{Discussion}
\label{sec_disc}

In this section we compare our results with other QCDSR calculations, in particular 
with  light-cone QCD sum rules (LCSR) results, and also with results obtained with 
other  techniques. Besides QCDSR, coupling constants can be estimated with the 
vector meson dominance (VMD) model, with heavy quark effective theory (HQET), with 
SU(4) symmetry relations,  with effective models such as the constituent quark-meson 
model, with chiral models and with lattice QCD calculations (LQCD).

\subsection{SU(4) and HQET}

If SU(4) would be exact, several  relations between the coupling 
constants should hold. Using the QCDSR results reported in Table \ref{tab2} we 
can check to what extent these relations are satisfied. The relevant relations 
and their deviation from our results  are shown in Table \ref{tab_comp}.   They are ordered by
 increasing degree of violation. Remembering 
the typical uncertainties of 10 \% to 15 \% in our calculations, we can 
conclude that the first seven  SU(4) relations are reasonably satisfied whereas 
the last six relations are badly violated. Although there is no rigorous  
systematics, we can clearly observe that violations occur mostly when there is a 
pion in the vertex. 
\begin{table}[htb]
\centering
\begin{tabular}{cc}
\hline
SU(4) Relation & Violation \\ \hline
$g_{J/\psi DD}=g_{J/\psi D^*D^*}$  & $(7 \mbox{\%})$  \\ 
$g_{\rho DD^*}=\frac{\sqrt{6}}{2}g_{J/\psi DD^*} $ & $(12 \mbox{\%})$  \\ 
$g_{\rho DD}=\frac{\sqrt{6}}{4}g_{J/\psi DD} $  & $ (17 \mbox{\%})$  \\ 
$g_{\pi D^*D^*}=\frac{\sqrt{6}}{2}g_{J/\psi DD^*}$  & $ (20 \mbox{\%})$  \\ 
$g_{D^* D^* \rho}= \frac{\sqrt{6}}{4} \, g_{J/\psi D^* D^*} $ & $(20 \mbox{\%})$ \\ 
$g_{D D \rho}= \frac{\sqrt{6}}{4} \, g_{J/\psi D^* D^*}$ & $(21 \mbox{\%})$ \\ 
$g_{\rho D^*D^*}=\frac{\sqrt{6}}{4}g_{J/\psi DD} $ & $(25 \mbox{\%})$ \\ 
$g_{\pi D^*D^*}=g_{\rho DD^*}$ &  $(29 \mbox{\%})$ \\ 
$g_{\rho DD}=g_{\rho D^*D^*}$  & $(36 \mbox{\%})$ \\ 
$g_{D^* D \pi}=g_{D^*  D^* \rho}$  & $(52 \mbox{\%})$ \\ 
$g_{D^* D \pi}= \frac{\sqrt{6}}{4} \, g_{J/\psi D^* D^*}$ & $(62 \mbox{\%})$ \\ 
$g_{D^* D \pi}= \frac{\sqrt{6}}{4} \, g_{J/\psi D D} $ & $(64 \mbox{\%})$ \\ 
$g_{D^* D \pi}=g_{D D \rho}$ &  $(70 \mbox{\%})$  \\ 
\hline
\end{tabular}
\caption{SU(4) relations between the coupling constants (on the left column) 
and their violation (in percentage on the right column) found in QCDSR.}
\label{tab_comp}
\end{table}

We can also check the heavy quark spin symmetry relations  \cite{dea}, which are 
presented below with their deviation from our results (calculated with the values 
presented in Table \ref{tab2} in parenthesis: 
\begin{eqnarray}
&&g_{\rho D^* D}={g_{\rho D^* D^*}\over m_{D^*}} \,\,\,\,\,\,\,\,\,\,\,\,\,\, (45 \mbox{\%}) 
\label{hqet1} \\
&&g_{J/\psi  D^* D}={g_{J/\psi D D}\over m_{D}} \,\,\,\,\,\,\,\, (22 \mbox{\%})
\label{hqet2}
\end{eqnarray}
These relations come from HQET.  
Since charm is not  heavy enough to ensure the validity of HQET, we would expect 
a significant violation of the above relations. In fact, considering the error in our 
calculations (\ref{hqet2}) is still satisfied whereas  (\ref{hqet1}) is severely 
violated.

\subsection{Light cone sum rules}

\begin{table}
\centering
\begin{tabular}{ccc}
\hline
Approach & $g_{D^*D\pi}$ &  $g_{B^*B\pi}$ \\ 
\hline
QCDSR \cite{col} & $9\pm2$ & 
$20\pm4$\\
QCDSR \cite{col} & $7\pm2$ & $15\pm4$\\ 
LCSR \cite{bel} & $11\pm2$ & $28\pm6$ \\
QCDSR \cite{dn} & $6.3\pm1.9$ & $14\pm4$\\ 
LCSR \cite{kho} & $10.5\pm3$ & $22\pm9$\\
QCDSR \cite{02r} & $14.0\pm 1.5$ & $42.5\pm2.6$\\
QCDSR plus meson loops \cite{mane} & $17.5\pm 1.5$ & $44.7\pm1.0$\\
LQCD \cite{Beci} & $20\pm 2$ & \\
LQCD \cite{Abada} & $18.8^{+2.5}_{-3.0}$& \\
dispersive quark model \cite{Melik} & $18\pm 3$ & $32\pm5$\\
Dyson-Schwinger equations \cite{bruno} & $15.8^{+2.1}_{-1.0}$ & 
$30.0^{+3.2}_{-1.4}$\\ 
\hline 
\end{tabular}
\caption{Summary of estimates for  $g_{D^*D\pi}$ and $g_{B^*B\pi}$. These couplings
refer to charged mesons $\pi^{\pm}$.}
\label{tabcompi}
\end{table}

In Table \ref{tabcompi} we present a compilation of the estimates of the 
coupling constants $g_{D^*D\pi}$ and $g_{B^*B\pi}$  from distinct  
calculations.
From this Table we see that our result in Ref.~\cite{02r} is in a 
fair agreement with the LCSR calculation in Refs.~\cite{bel,kho}, but is 
still smaller than the
experimental value \cite{cleo}: $g_{D^*D\pi}=17.9\pm0.3\pm1.9$. However, 
using a better way to extrapolate the QCDSR results based on a meson
loop calculation, the result obtained in Ref.~\cite{mane} is
in an excellent agreement with the experimental value. From this Table
we also see that LQCD results are in a very good agreement with the 
experimental value.

The basic difference between the QCDSR, described here, and the LCSR is the 
fact that, instead of considering the three-point function in 
Eq.~(\ref{corr}),
the central object in the LCSR is the correlation function of two meson 
currents between the vacuum and one on-shell meson state \cite{cokho}:
\beq
\Gamma (q,p) = \int d^4x \;\; e^{iq\cdot x} 
\,  \langle M_1(p)|T\{j_{3}(x) j_{2}^{\dagger}(0)\}|0\rangle. 
\label{corlc} 
\enq
The idea is to expand the product of the currents near the light-cone
$x^2=0$. This expansion is different from the local OPE expansion used in the
QCDSR because it incorporates a summation of an infinite series of local 
operators \cite{cokho}. In the case that $M_1$ is the pion, the vacuum-pion
matrix elements are expressed via pion light-cone distribution amplitudes 
(DA). Since the pion DA's have a well defined twist, the LCSR can be written 
in terms of a twist expansion. On  the other hand, if $M_1$ is not light,
its DA does  not have a well defined twist. If $M_1$ is a heavy meson, like a 
$B$ meson, the correlation function can be systematically expanded in the
limit of large $m_b$ in heavy quark effective theory \cite{kho2}. However, 
for hadronic  vertices involving only charmed mesons there is no  well defined 
expansion for the correlation function in the LCSR approach. An attempt to 
use LCSR to calculate the $D^* D^*  \rho$ coupling constant was performed in 
\cite{zhizhi}. The obtained value is quoted in Table \ref{tabvec}. It is closer  
to SU(4) and to VMD (see below) than the value found in \cite{08}.  However, a  
discussion of  some potential sources of uncertainties, such as the use of a 
relatively large value of the Borel mass and the possible  continuum dominance,  
is still missing in \cite{zhizhi}.

\subsection{Vector meson dominance}

In the VMD model \cite{vmd} a virtual photon with four-momentum 
$q$ is emitted in the process $e M\to e M$, where $M$ represents 
any meson. The photon  can be decomposed into a sum of all neutral vector mesons 
including both isospin 0 and isospin 1. Then the vector meson couples with the
external  meson $M$. At $q^2 = 0$  one can write:
\beq
\sum_{V=\rho,\omega,\phi,J/\psi,...}{\gamma_V g_{VMM}\over m_V^2}=e,
\label{gvmd}
\enq 
where $\gamma_V$ is the photon-vector-meson  coupling that can be 
determined from the vector meson partial decay width to $e^+e^-$:
\beq
\Gamma_{V\to ee}={\alpha\gamma_V^2\over3m_V^3}.
\enq

Using the VMD model the couplings $g_{\rho DD},~g_{\rho D^*D^*},~g_{J/\psi
 DD}$ and $g_{J/\psi D^* D^*}$ were estimated in Ref.\cite{ko}.
The couplings $g_{\rho D^*D}$ and $g_{J/\psi D^* D}$  were estimated in 
Ref. \cite{osl} by applying the VMD model to the radiative decay 
$D^*\to D\gamma$. The obtained values are presented in the 
Table~\ref{tabvec} where  we also present a summary of the predictions for
the coupling constants in the $VD^{(*)}D^{(*)}$ vertex.

In our approach the coupling constant is given by the value of the form factor 
at $Q^2 = -m^2_M$, where $M$ is the off-shell vector meson. 
In the case of $J/\psi$ this corresponds 
to  $Q^2 = - 9.6$ GeV$^2$ whereas for $\rho_0$ the on-shell point is 
$Q^2 = - 0.6$ GeV$^2$. In the VMD model the vector meson has $Q^2 = 0$ GeV$^2$. 
We would then expect  our results to  present a reasonable agreement with the 
VMD estimates of the  $\rho$ couplings and a significant disagreement with the 
VMD estimates of the  $J/\psi$ couplings. This  pattern is not so evident 
in Table  \ref{tabvec}, which shows the  approximate expected behavior in some cases   
but also shows discrepant behavior in other cases.  
All the form factors decrease as $Q^2$ goes from the time-like to the 
space-like region and hence all the values shown in the left column of  Table  
\ref{tabvec} will become smaller when calculated at $Q^2 = 0$ GeV$^2$. This will improve 
the agreement with the VMD estimates in the case of the light vector mesons and will 
increase the disagreement with the heavy vector mesons. Since QCDSR are more reliable in 
the latter case, this comparison suggests that the use of VMD for heavy mesons is 
dangerous.  

\begin{table}
\centering
\begin{tabular}{cccc}
\hline
Coupling & QCDSR & VMD & Other models \\
\hline
$g_{\rho DD}$  & $3.0\pm0.2$ \cite{01} & 2.52 \cite{ko} & \\
$g_{\rho D^*D}$ (GeV$^{-1}$)  & $4.3\pm0.9$ \cite{10} & 2.82 
\cite{osl} & $4.17\pm1.04$ \cite{dai02}\\ 
$g_{\rho D^*D^*}$   & $4.7\pm0.2$ \cite{08} & 2.52 
\cite{ko} & $1.8 \pm 0.5$ \cite{zhizhi} \\
$g_{\omega DD}$  & $-2.9$ \cite{mihara} & -2.84 \cite{ko} & \\
$g_{J/\psi DD}$  & $5.8\pm0.9$ \cite{05i} & 7.64 \cite{ko} & 
$8.0\pm0.5$ \cite{dea} \\
$g_{J/\psi D^*D}$ (GeV$^{-1}$)   & $4.0\pm0.6$ \cite{05i} & $8.0\pm0.6$ 
\cite{osl} & $4.05\pm0.25$ \cite{dea} \\ 
$g_{J/\psi D^*D^*}$  & $6.2\pm0.9$ \cite{05} & 7.64 \cite{ko} & $
8.0\pm0.5$ \cite{dea}\\
\hline 
\end{tabular}
\caption{Summary of estimates for  $g_{VD^{(*)}D^{(*)}}$. In the above couplings
$\rho$ stands  for $\rho_0$.}
\label{tabvec}
\end{table}

\section{ $J/\psi$  absorption and production}
\label{sec_psi}

\begin{figure}
\begin{center}
\epsfxsize=10cm
\leavevmode
\hbox{\epsffile{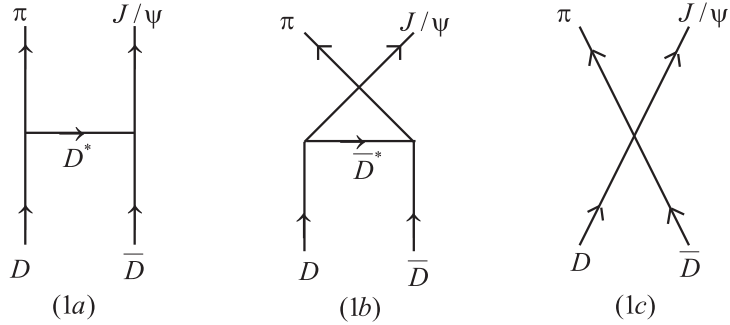}}
\end{center}
\protect\caption{Diagrams which contribute to the process 
$D \bar{D}\rightarrow J/\psi+\pi$.}
\label{psiprod1}
\end{figure}

\begin{figure}
\begin{center}
\epsfxsize=10cm
\leavevmode
\hbox{\epsffile{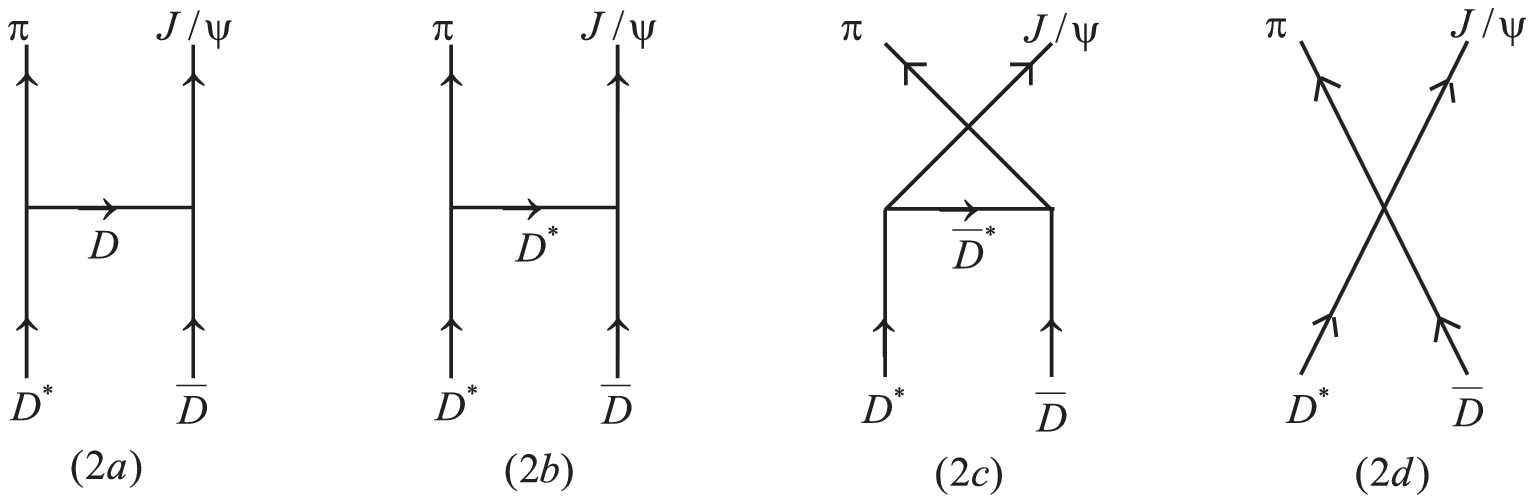}}
\end{center}
\protect\caption{Diagrams which contribute to the process
$D^* \bar{D}\,\rightarrow J/\psi + \pi$.}
\label{psiprod2}
\end{figure}

\begin{figure}
\begin{center}
\epsfxsize=10cm
\leavevmode
\hbox{\epsffile{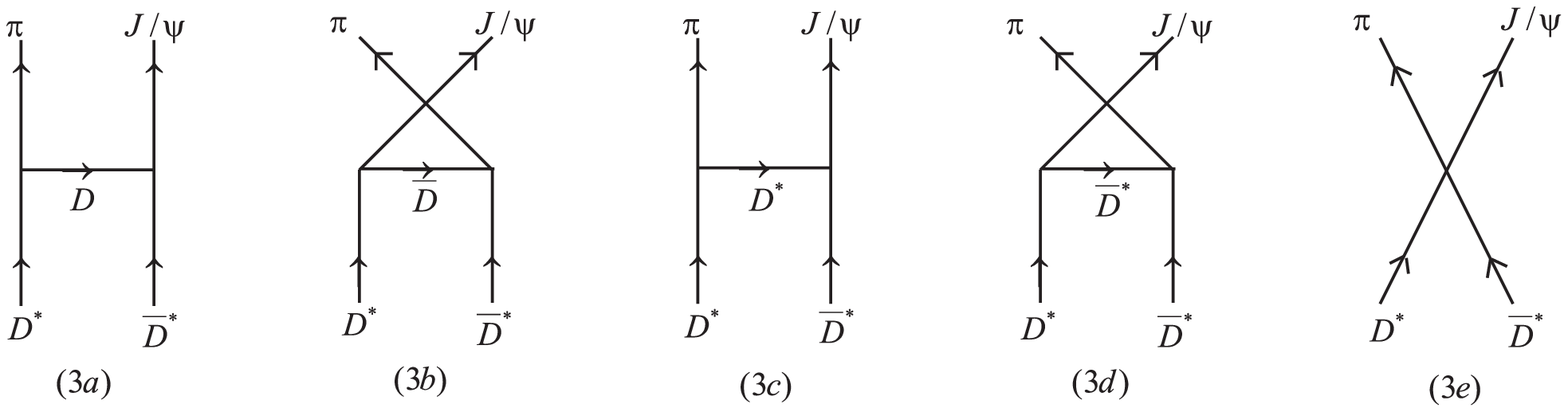}}
\end{center}
\protect\caption{ Diagrams which contribute to the process
$D^* \bar{D^*}\,\rightarrow J/\psi + \pi$.}
\label{psiprod3}
\end{figure}

As an application of the form factors obtained above we address now the 
problem of $J/\psi$ absorption and production in hadronic matter. 
With the Lagrangians (\ref{dsdpi}) -  (\ref{psidsdspi}) 
we are able to compute the process 
$D \bar{D}\rightarrow J/\psi+\pi$, which involves the diagrams in
Fig. \ref{psiprod1}, the process $D^* \bar{D}\,\rightarrow J/\psi + \pi$,   
corresponding to the diagrams shown in Fig. \ref{psiprod2}  and also the process
$D^* \bar{D^*}\,\rightarrow J/\psi + \pi$, corresponding to the diagrams 
in Fig. \ref{psiprod3}.

As extensively discussed in previous works, although the above Lagrangians and 
amplitudes are quite satisfactory from the point of view of symmetry requirements, 
their straightforward application to the computation of cross sections leads to 
unacceptably large results. This comes from the fact that the exchanged particles may 
be far off-shell and therefore they enter (or leave) a vertex with a very  different 
resolving power. In one extreme case, a virtual $J/\psi$ probing a $D$ meson, may behave 
like a parton. Of course, when this happens, the compact $J/\psi$ almost misses the 
large $D$ and  as a consequence the cross section of the whole process drops 
significantly. This physics of  spatial extension and resolving power is contained in the 
form factors. It has been realized by many authors that calculations with and without form 
factors lead to results differing by up to two orders of magnitude! Therefore we simply 
{\it can not ignore the form factors}. We must include them in order to obtain reliable 
results! 

Looking at the diagrams in Figs. \ref{psiprod1}, \ref{psiprod2}  and   \ref{psiprod3} 
we notice that we need the following form factors (and the corresponding coupling constants):
$g^{(D^{*})}_{\pi D D^{*}}(t)$, 
$g^{(D)}_{J/\psi D D}(t)$, 
$g^{(D^{*})}_{J/\psi D D^{*}}(t)$, 
$g^{(D)}_{J/\psi DD^{*}}(t)$, 
$g^{(D^{*})}_{J/\psi D^{*} D^{*}}(t)$ and 
$g^{(D^{*})}_{\pi D^{*} D^{*}}(t) $, 
where $t$ is the usual momentum transfer squared and in the superscript in 
parenthesis we denote the off-shell particle. This is an important distinction, 
because the form factors in the same vertex, as we have seen, are very different when 
different particles  are off-shell.

\begin{figure}
\begin{center}
\epsfxsize=9cm
\leavevmode
\hbox{\epsffile{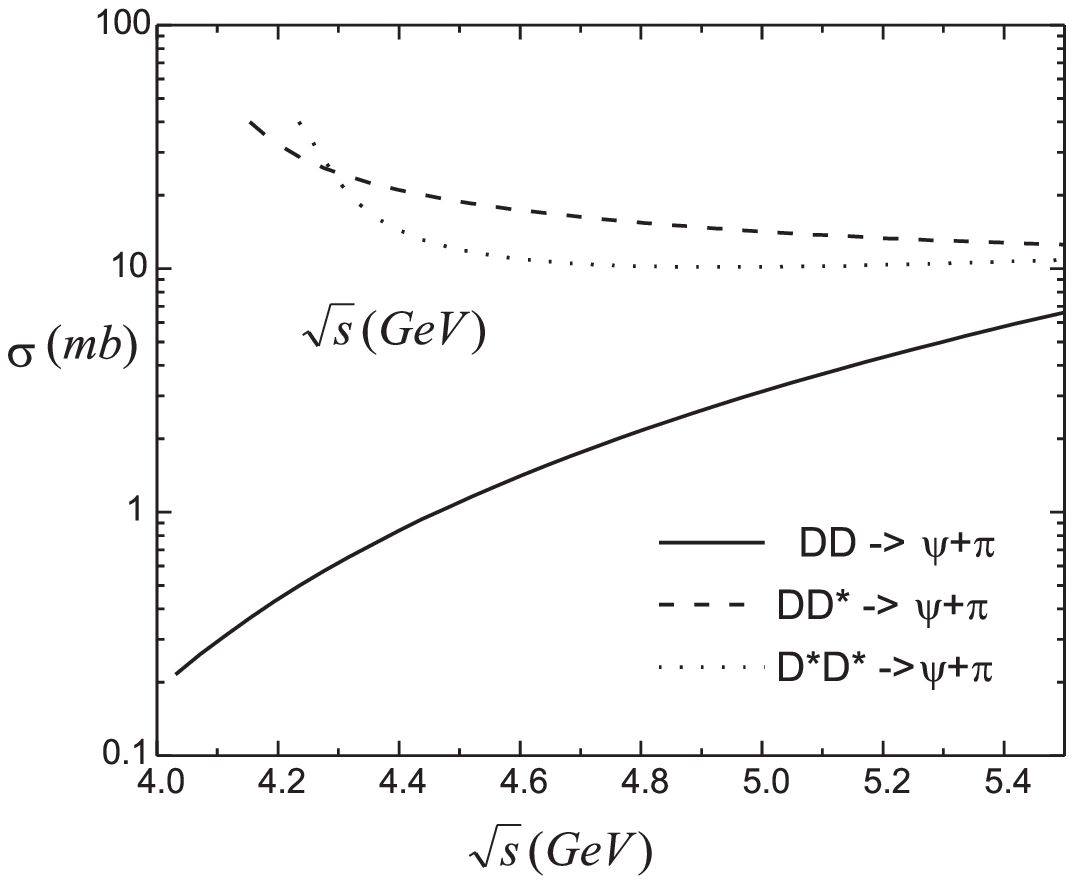}}
\end{center}
\protect\caption{$J/\psi$ secondary production cross section without form factors.}
\label{psiprod6}
\end{figure}

\begin{figure}
\begin{center}
\epsfxsize=9cm
\leavevmode
\hbox{\epsffile{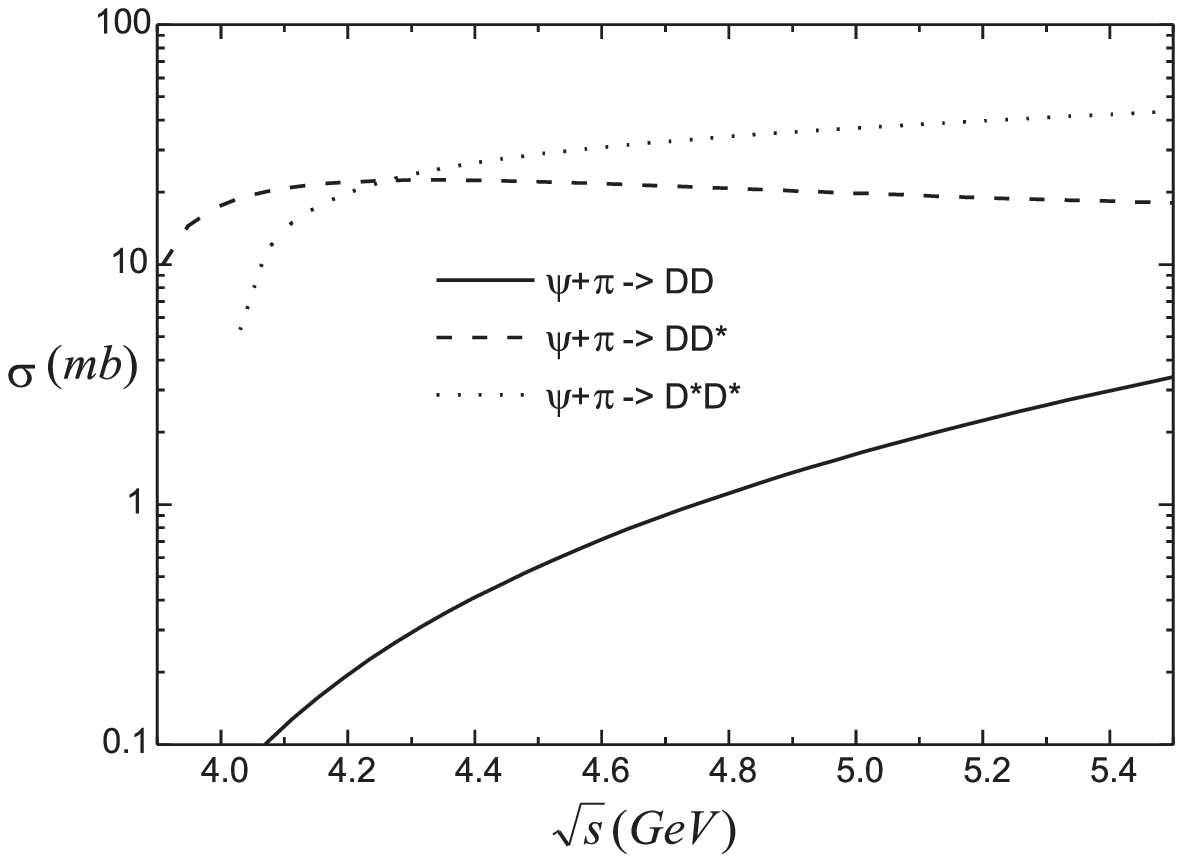}}
\end{center}
\protect\caption{$J/\psi$ absorption cross section obtained through detailed balance without
form factors.}
\label{psiprod7}
\end{figure}

\begin{figure}
\begin{center}
\epsfxsize=9cm
\leavevmode
\hbox{\epsffile{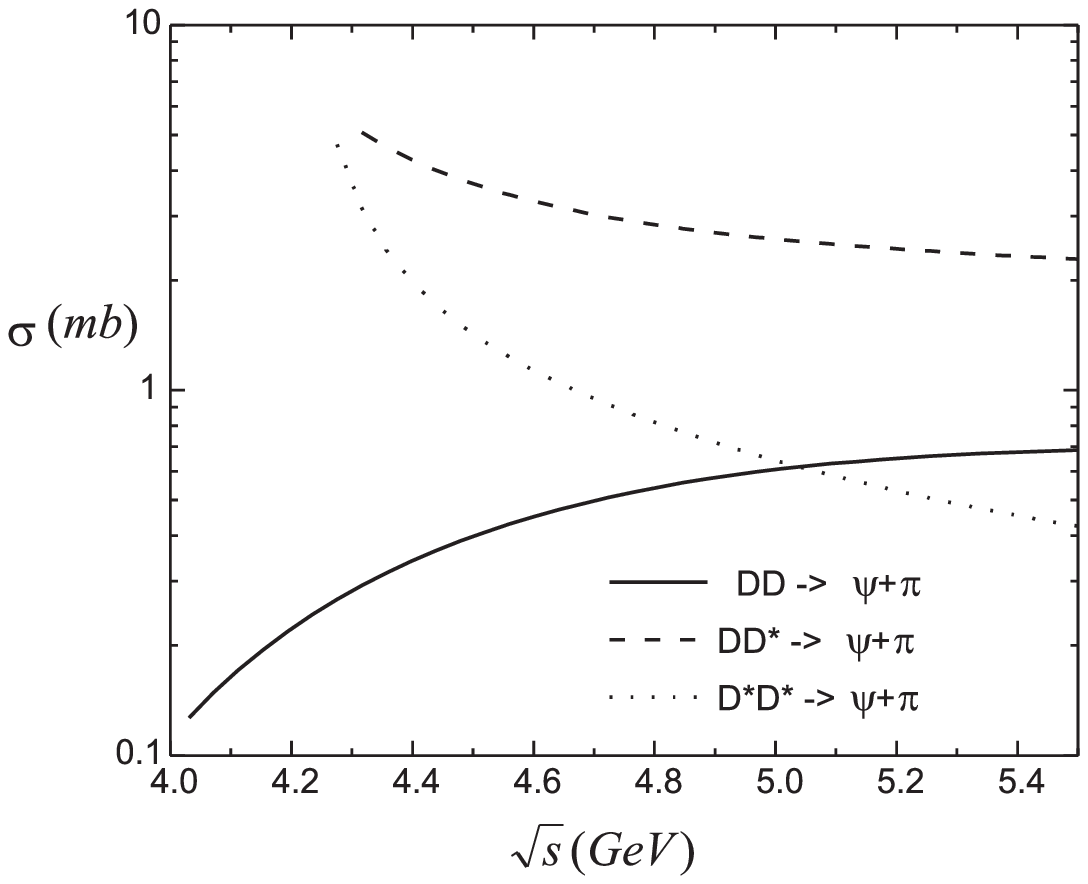}}
\end{center}
\protect\caption{$J/\psi$ secondary production cross section with form factors.}
\label{psiprod8}
\end{figure}

\begin{figure}
\begin{center}
\epsfxsize=9cm
\leavevmode
\hbox{\epsffile{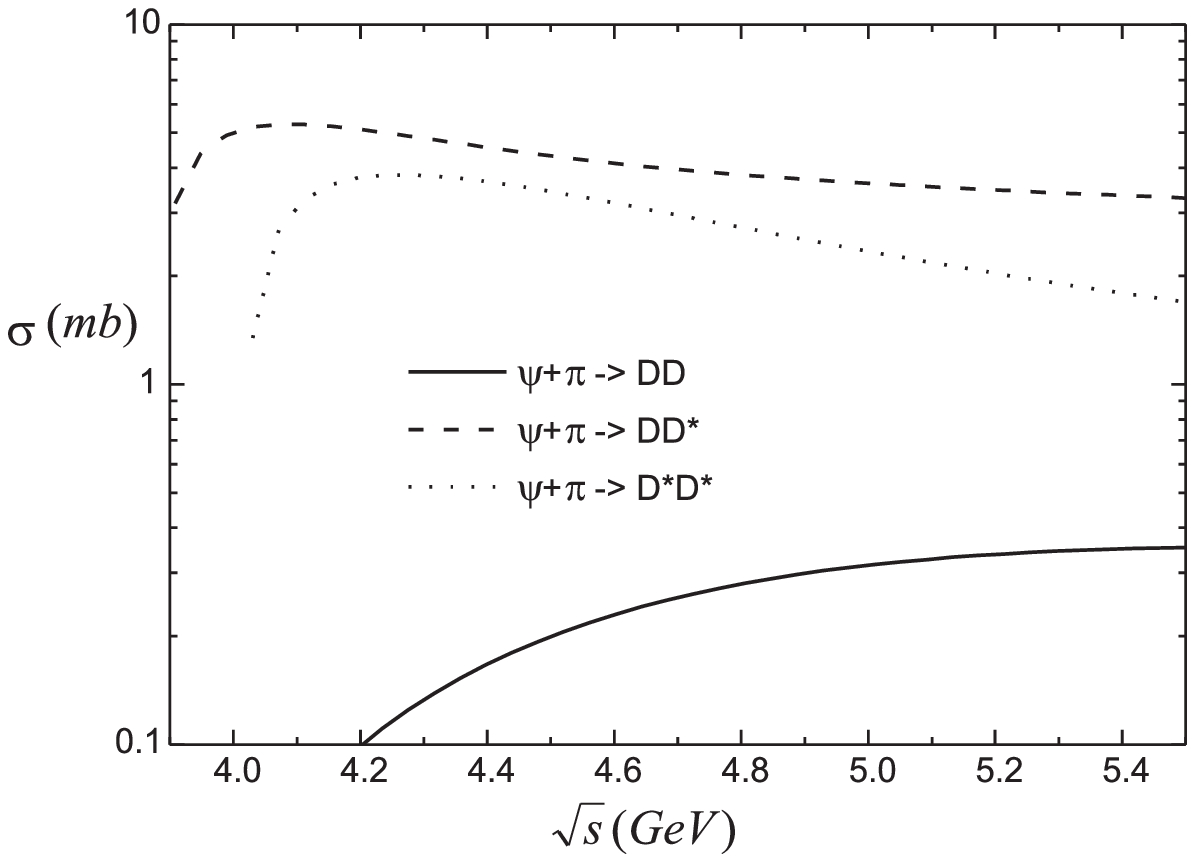}}
\end{center}
\protect\caption{$J/\psi$ absorption cross section obtained through detailed balance with
form factors.}
\label{psiprod9}
\end{figure}

The cross sections for secondary $J/\psi$ production is  related to  the 
annihilation through detailed balance. 
In Fig. \ref{psiprod6} we show the $J/\psi$ secondary production cross section as a 
function of $\sqrt{s}$, without form factors. In all Figures, the channels 
$D \bar{D}\rightarrow J/\psi+\pi$, $D \bar{D^*}\rightarrow J/\psi+\pi$ and 
$D^* \bar{D^*}\rightarrow J/\psi+\pi$ are represented by solid, dashed and dotted 
lines respectively. In Fig. \ref{psiprod7}  we show the 
corresponding inverse reactions. As it can be seen, the cross sections have the 
same order of magnitude in both directions. Figs.  \ref{psiprod8}  and 
\ref{psiprod9} are the analogues of \ref{psiprod6} and   \ref{psiprod7}  when we 
include the form factors in the calculations.  Of course, only these last two Figures 
correspond to realistic numbers. The comparison of the two sets of Figures is interesting  
to estimate the effect of form factors. In previous studies doing the same kind of 
comparison, as for example in \cite{osl}, the introduction of form factors reduced the cross 
sections 
by factors ranging between  $20$  and  $50$ depending on the channel. In that work 
the form factor was the same for all 
vertices and the cut-off,  not known,   was estimated to be between $1$ and $2$ GeV. 
Our study is much more detailed and not only each vertex has its own form factor,
but, depending on which particle is off-shell the form factor is different. The final
effect of all these peculiarities is the reduction of the cross sections by a factor
around $7$. Although significant, this reduction is smaller than previously expected.

Fig. \ref{psiprod8} contains our main results. 
The plotted cross sections can be compared with 
the results of \cite{ko98} and, more directly,  with \cite{brat}. In Fig.  2 of 
\cite{ko98}, although the variables in the plot are different, we can observe the same 
trend and relative importance of the three channels. In that work, the results were 
obtained 
with the quark model of  \cite{mbq}. Our curves share some features with the results of 
\cite{brat}, such as, for example, the dominance of the $D D^*$ channel and the falling 
trend of the $DD^*$ and $D^* D^*$ channels.  The behavior of the $DD$ channel is quite 
different. In the energy range of $\sqrt{s} > 4.5$ GeV our cross sections are smaller by a
factor of $2$ ($DD^*$) or $5$ ($D^* D^*$ and $D D$). These discrepancies are large but they 
are expected  since in \cite{brat} all channels include the final state $J/\psi + \rho$, 
which we did not include. In the model used by the Giessen group \cite{brat}  the cross 
sections for $D + \bar{D}  \rightarrow J/\psi + \pi$ and 
$D + \bar{D} \rightarrow  J/\psi + \rho$ are 
similar and the same conclusion holds for the other inital state open charm mesons.  If
this would remain true  in the effective Lagrangian approach, then our results 
including both final states would come closer to those of \cite{brat}, giving thus a more 
theoretical support to the model considered there. 

The exercise presented in this section was meant to illustrate the use  
of the charm form factors discussed in this review. The ultimate computation of charmonium 
interactions in a hot and dense medium should include  other effects, not mentioned here. 
For recent works on the subject see, for example, \cite{bourque,crochet}.

\section{Summary}
\label{sec_conc}
\nin
We have studied the form factors of vertices with charm mesons. They are relevant to understand 
data from  heavy ion collisions at RHIC and LHC, from B decays at 
BELLE and BABAR and, in the future, charm production at PANDA. We have described how to 
calculate them with QCDSR and presented the results of our calculations.  The comparison of the 
obtained coupling constants  with those obtained with other methods shows that  all the numbers have the 
same order of magnitude and the discrepancies between them go from a few percent to a factor two.  In 
every approach there are improvements to be made and more accurate results are expected in the future. 
There are still vertices which have not been studied, such as those with $\eta_c$. With the already 
available charm form factors it is possible to address a number of problems of phenomenological interest. 
One of them, namely, the production and absorption of $J/\psi$ in nuclear matter was discussed here and 
the huge effect of form factors was shown in detail.  Moreover some of these form factors   have been
used in \cite{molina1} to study resonances formed through meson exchange interactions.

Most of the calculations 
follow  standard procedures in QCDSR but the extrapolation techniques to compute the coupling constant were  
developed by our group. We calculated two form factors for two off-shell mesons  in the considered vertex. 
The simultaneous extrapolation of these form factors allowed the determination of the coupling constant. 
While this requirement  proved to be crucial for finding the coupling, the inclusion of a third form factor,  
obtained by putting the third meson off the mass shell, with the subsequent triple extrapolation to obtain 
the coupling, did not bring any significant improvement in the  results.  

The extrapolation method used in our works 
has a systematic error which comes from the choice of the analytic form of the extrapolating functions. We 
considered only monopole, exponential and gaussian parametrizations. However there is no physical reason for 
choosing these forms. Using the $D^* D \pi$ vertex for an exploratory study, we calculated the form 
factor computing the relevant hadronic loops. In this way there is no need to guess a particular form for the
form factor. Since the hadronic loop calculation contains non-perturbative physics it should be reliable 
in low $Q^2$ domain.   The introduction of hadronic loops improved a lot our quantitative analysis of the  
$D^* D \pi$ vertex. However we think that it would be premature to include it as an obligatory complement to  the 
QCDSR formalism. Some loop diagrams were neglected because they were assumed to be less important but this must still 
be proven. Moreover the effective Lagrangian theories for other mesons and for baryons are less well known. 

The disagreement between some of our results and the SU(4)  predictions has still to be clarified. The same is true 
for the estimates made with the help of light cone sum rules, heavy quark symmetry 
and vector meson dominance. From the perspective of QCDSR there is still room for improvements, as, for 
example, the inclusion of $\alpha_s$ corrections and higher order condensates.   
An important extension of our program will be to systematically study 
vertices with  charm and strange mesons.  Some of them were already considered in 
Refs.~\cite{mcc,as,sssy,wang1,wang2}.

Beyond individual technical improvements, we believe that the QCDSR community should  make a joint effort dedicated to 
systematically compare results and methods and arrive at a consensus on the present status of the calculation of 
coupling constants. With this review we wish to take a step in this direction.

\addcontentsline{toc}{section}{Acknowledgments}
\section*{Acknowledgments}

This work has been partly supported by the brazilian funding agencies FAPESP and
CNPq. We are deeply grateful to F. Carvalho, F.O. Dur\~aes, A. Lozea,  R.D. Matheus,  
B. Os\'orio Rodrigues, M.R. Robilotta, R. Rodrigues da Silva, C. Schat, S Narison, 
Q. Zhao, E. Oset and  E. Shuryak, for fruitful discussions.






\end{document}